\documentclass{aastex61}
\usepackage[caption=false]{subfig}
\shortauthors{Oelkers et al. 2025}

\begin{document}

\title{Ground-Based Reconnaissance Observations of 21 Exoplanet Atmospheres \\
with the Exoplanet Transmission Spectroscopy Imager}

\correspondingauthor{Ryan J. Oelkers}
\email{ryan.oelkers@utrgv.edu}

\author{Ryan J. Oelkers}
\affil{Department of Physics and Astronomy, The University of Texas, Rio Grande Valley, Brownsville, TX 78520, USA}
\affil{Department of Physics and Astronomy, Texas A\&M University, College Station, TX, 77843-4242 USA}
\affil{George P. and Cynthia Woods Mitchell Institute for Fundamental Physics and Astronomy, Texas A\&M University, College Station, TX, 77843-4242 USA}
\affil{Charles R. \& Judith G. Munnerlyn Astronomical Laboratory, Department of Physics \& Astronomy, Texas A\&M University, College Station, TX 77843, USA}

\author{Luke M. Schmidt}
\affil{Department of Physics and Astronomy, Texas A\&M University, College Station, TX, 77843-4242 USA}
\affil{George P. and Cynthia Woods Mitchell Institute for Fundamental Physics and Astronomy, Texas A\&M University, College Station, TX, 77843-4242 USA}
\affil{Charles R. \& Judith G. Munnerlyn Astronomical Laboratory, Department of Physics \& Astronomy, Texas A\&M University, College Station, TX 77843, USA}

\author{Erika Cook}
\affil{Department of Physics and Astronomy, Texas A\&M University, College Station, TX, 77843-4242 USA}
\affil{George P. and Cynthia Woods Mitchell Institute for Fundamental Physics and Astronomy, Texas A\&M University, College Station, TX, 77843-4242 USA}
\affil{Charles R. \& Judith G. Munnerlyn Astronomical Laboratory, Department of Physics \& Astronomy, Texas A\&M University, College Station, TX 77843, USA}

\author{Mary Anne Limbach}
\affil{Department of Astronomy,University of Michigan, Ann Arbor, MI 48109 USA}
\affil{Department of Physics and Astronomy, Texas A\&M University, College Station, TX, 77843-4242 USA}
\affil{George P. and Cynthia Woods Mitchell Institute for Fundamental Physics and Astronomy, Texas A\&M University, College Station, TX, 77843-4242 USA}
\affil{Charles R. \& Judith G. Munnerlyn Astronomical Laboratory, Department of Physics \& Astronomy, Texas A\&M University, College Station, TX 77843, USA}

\author{D. L. DePoy}
\affil{Department of Physics and Astronomy, Texas A\&M University, College Station, TX, 77843-4242 USA}
\affil{George P. and Cynthia Woods Mitchell Institute for Fundamental Physics and Astronomy, Texas A\&M University, College Station, TX, 77843-4242 USA}
\affil{Charles R. \& Judith G. Munnerlyn Astronomical Laboratory, Department of Physics \& Astronomy, Texas A\&M University, College Station, TX 77843, USA}

\author{J. L. Marshall}
\affil{Department of Physics and Astronomy, Texas A\&M University, College Station, TX, 77843-4242 USA}
\affil{George P. and Cynthia Woods Mitchell Institute for Fundamental Physics and Astronomy, Texas A\&M University, College Station, TX, 77843-4242 USA}
\affil{Charles R. \& Judith G. Munnerlyn Astronomical Laboratory, Department of Physics \& Astronomy, Texas A\&M University, College Station, TX 77843, USA}

\author{Jimmy Ardoin}
\affil{Department of Astronomy, The University of Texas at Austin, Austin, TX 78712, USA}
\affil{Department of Physics and Astronomy, Texas A\&M University, College Station, TX, 77843-4242 USA}
\affil{Charles R. \& Judith G. Munnerlyn Astronomical Laboratory, Department of Physics \& Astronomy, Texas A\&M University, College Station, TX 77843, USA}

\author{Mitchell Barry}
\affil{Department of Physics and Astronomy, Texas A\&M University, College Station, TX, 77843-4242 USA}
\affil{Charles R. \& Judith G. Munnerlyn Astronomical Laboratory, Department of Physics \& Astronomy, Texas A\&M University, College Station, TX 77843, USA}

\author{Evan Batteas}
\affil{Department of Physics and Astronomy, Texas A\&M University, College Station, TX, 77843-4242 USA}
\affil{Charles R. \& Judith G. Munnerlyn Astronomical Laboratory, Department of Physics \& Astronomy, Texas A\&M University, College Station, TX 77843, USA}

\author{Alexandra Boone}
\affil{Department of Physics \& Astronomy, University of Wyoming, Laramie, WY 82071 USA}
\affil{Department of Physics and Astronomy, Texas A\&M University, College Station, TX, 77843-4242 USA}
\affil{Charles R. \& Judith G. Munnerlyn Astronomical Laboratory, Department of Physics \& Astronomy, Texas A\&M University, College Station, TX 77843, USA}

\author{Brant Conway}
\affil{Department of Physics and Astronomy, Texas A\&M University, College Station, TX, 77843-4242 USA}
\affil{Charles R. \& Judith G. Munnerlyn Astronomical Laboratory, Department of Physics \& Astronomy, Texas A\&M University, College Station, TX 77843, USA}

\author{Silvana Delgado Adrande}
\affil{Department of Physics and Astronomy, Texas A\&M University, College Station, TX, 77843-4242 USA}
\affil{George P. and Cynthia Woods Mitchell Institute for Fundamental Physics and Astronomy, Texas A\&M University, College Station, TX, 77843-4242 USA}

\author{John D. Dixon}
\affil{Department of Physics and Astronomy, Texas A\&M University, College Station, TX, 77843-4242 USA}
\affil{George P. and Cynthia Woods Mitchell Institute for Fundamental Physics and Astronomy, Texas A\&M University, College Station, TX, 77843-4242 USA}

\author{Enrique Gonzalez-Vega}
\affil{Department of Physics and Astronomy, Texas A\&M University, College Station, TX, 77843-4242 USA}
\affil{Charles R. \& Judith G. Munnerlyn Astronomical Laboratory, Department of Physics \& Astronomy, Texas A\&M University, College Station, TX 77843, USA}

\author{Alexandra Guajardo}
\affil{Department of Atmospheric Sciences, Texas A\&M University, College Station, TX 77843}
\affil{Charles R. \& Judith G. Munnerlyn Astronomical Laboratory, Department of Physics \& Astronomy, Texas A\&M University, College Station, TX 77843, USA}

\author{Landon Holcomb}
\affil{Department of Physics and Astronomy, Texas A\&M University, College Station, TX, 77843-4242 USA}
\affil{Charles R. \& Judith G. Munnerlyn Astronomical Laboratory, Department of Physics \& Astronomy, Texas A\&M University, College Station, TX 77843, USA}

\author{Christian Lambert}
\affil{Department of Physics and Astronomy, Texas A\&M University, College Station, TX, 77843-4242 USA}
\affil{Charles R. \& Judith G. Munnerlyn Astronomical Laboratory, Department of Physics \& Astronomy, Texas A\&M University, College Station, TX 77843, USA}

\author{Shravan Menon}
\affil{Department of Physics, Washington University in St. Louis, St. Louis, MO 63130-4899 USA}
\affil{Department of Physics and Astronomy, Texas A\&M University, College Station, TX, 77843-4242 USA}
\affil{Charles R. \& Judith G. Munnerlyn Astronomical Laboratory, Department of Physics \& Astronomy, Texas A\&M University, College Station, TX 77843, USA}

\author{Divya Mishra}
\affil{Department of Physics and Astronomy, Texas A\&M University, College Station, TX, 77843-4242 USA}
\affil{George P. and Cynthia Woods Mitchell Institute for Fundamental Physics and Astronomy, Texas A\&M University, College Station, TX, 77843-4242 USA}

\author{Jacob Purcell}
\affil{Department of Physics, Indiana University - Purdue University at Indianapolis (IUPUI), 402 North Blackford Street, Indianapolis, Indiana 46202-3273}
\affil{Department of Physics and Astronomy, Texas A\&M University, College Station, TX, 77843-4242 USA}
\affil{Charles R. \& Judith G. Munnerlyn Astronomical Laboratory, Department of Physics \& Astronomy, Texas A\&M University, College Station, TX 77843, USA}

\author{Zachary Reed}
\affil{Department of Physics and Astronomy, Texas A\&M University, College Station, TX, 77843-4242 USA}
\affil{Charles R. \& Judith G. Munnerlyn Astronomical Laboratory, Department of Physics \& Astronomy, Texas A\&M University, College Station, TX 77843, USA}

\author{Nathan Sala}
\affil{Department of Physics and Astronomy, Texas A\&M University, College Station, TX, 77843-4242 USA}
\affil{Charles R. \& Judith G. Munnerlyn Astronomical Laboratory, Department of Physics \& Astronomy, Texas A\&M University, College Station, TX 77843, USA}

\author{Noah Siebersma}
\affil{Department of Physics and Astronomy, Texas A\&M University, College Station, TX, 77843-4242 USA}
\affil{Charles R. \& Judith G. Munnerlyn Astronomical Laboratory, Department of Physics \& Astronomy, Texas A\&M University, College Station, TX 77843, USA}

\author{Nhu Ngoc Ton}
\affil{Department of Mathematics, Texas A\&M University, College Station, TX 77843}
\affil{Charles R. \& Judith G. Munnerlyn Astronomical Laboratory, Department of Physics \& Astronomy, Texas A\&M University, College Station, TX 77843, USA}

\author{Raenessa M. L. Walker}
\affil{Department of Physics and Astronomy, Texas A\&M University, College Station, TX, 77843-4242 USA}
\affil{George P. and Cynthia Woods Mitchell Institute for Fundamental Physics and Astronomy, Texas A\&M University, College Station, TX, 77843-4242 USA}

\author{Z. Franklin Wang}
\affil{Department of Physics and Astronomy, Texas A\&M University, College Station, TX, 77843-4242 USA}
\affil{George P. and Cynthia Woods Mitchell Institute for Fundamental Physics and Astronomy, Texas A\&M University, College Station, TX, 77843-4242 USA}
\affil{Charles R. \& Judith G. Munnerlyn Astronomical Laboratory, Department of Physics \& Astronomy, Texas A\&M University, College Station, TX 77843, USA}

\author{Kaitlin Webber}
\affil{Department of Physics and Astronomy, Texas A\&M University, College Station, TX, 77843-4242 USA}
\affil{George P. and Cynthia Woods Mitchell Institute for Fundamental Physics and Astronomy, Texas A\&M University, College Station, TX, 77843-4242 USA}

\begin{abstract}
One of the most prolific methods of studying exoplanet atmospheres is transmission spectroscopy, which measures the difference between the depth of an exoplanet's transit signal at various wavelengths and attempts to correlate the depth changes to potential features in the exoplanet's atmosphere. Here we present reconnaissance observations of 21 exoplanet atmospheres measured with the Exoplanet Transmission Spectroscopy Imager (ETSI), a recently deployed spectro-photometer on the McDonald Observatory Otto Struve 2.1~m telescope. ETSI measurements are mostly free of systematics through the use of a novel observing technique called common-path multi-band imaging (CMI), which has been shown to achieve photometric color precision on-par with space-based observations (300ppm or 0.03\%). This work also describes the various statistical tests performed on the data to evaluate the efficacy of the CMI method and the ETSI instrument in combination. We find that none of the 8 comparisons of exoplanet atmospheres measured with ETSI and other observatories (including the Hubble Space Telescope) provide evidence that the spectra are statistically dissimilar. These results suggest that ETSI can provide initial transmission spectroscopy observations for a fraction of the observational and monetary overhead previously required to detect an exoplanet's atmosphere. Ultimately these reconnaissance observations increase the number of planets with transmission spectroscopy measurements by $\sim10\%$ and provide an immediate prioritization of 21 exoplanets for future follow-up with more precious observatories, such as the James Webb Space Telescope. The reconnaissance spectra are available through the Filtergraph visualization portal at the URL \url{https://filtergraph.com/etsi/}.
\end{abstract}

\section{Introduction} \label{sec:intro}

Characterizing the atmospheres of gas giant exoplanets provides several insights into exoplanet formation theory. First, these measurements supply information about the physical and chemical processes occurring in the exoplanet's present-day atmosphere (via the detection of condensation clouds and photochemical hazes), which ultimately provide information about the planet's composition and can provide details about the formation and evolutionary history of the planet. Second, ultra-precise atmospheric observations can differentiate whether a given planet has a solid or gaseous surface near planetary radius-mass boundaries. Finally, a detailed study of the molecular make-up of a given atmosphere contributes to the understanding of atmospheric processes across planet types and has the potential to impart knowledge about the habitability of other worlds. 

One method used to study exoplanet atmospheres is transmission spectroscopy, which measures the difference between the depth of an exoplanet's transit signal at various wavelengths and attempts to correlate the depth changes to potential features in the exoplanet's atmosphere \citep{Seager:2000, Kreidberg:2018}. Observations using transmission spectroscopy have already successfully detected atomic and molecular absorption features in approximately 100 exoplanet atmospheres and it is one of the most productive methods of studying atmospheres to date \citep{sing:2008, pont:2013, Sing:2011, Wilson:2015, Sing:2016, Turner:2016, Alam:2018, Alam:2020, Feinstein:2022, Fu:2022, Rustamkulov:2022}. 

Recent ensemble studies of exoplanet transmission spectra suggest a diversity of molecular features exist in exoplanet atmospheres. However, these ensemble studies have shown very little evidence of verifiable trends in atmospheric structure and/or make-up and the evolutionary nature of gas giant planets is still ambiguous \citep{Dymont:2022, Mansfield:2022, Edwards:2023}. Similarly, recent stellar variability studies indicate stellar contamination can alter measured atmospheric abundances and temperature by several orders of magnitude \citep{Saba:2024}. Additionally, a number of planetary structure models poorly account for all variations of known exoplanets and include predictions for objects which appear to lie in empirical exoplanet ``deserts" \citep{Kirk:2022, Thorngren:2022}. These results indicate exoplanet formation may be non-homogeneous and different exoplanets may require completely different formation mechanisms. It remains to be fully investigated whether these missing trends across the population are due to selection effects in the sample, environmental effects from the host star's variability, weather effects on the planets during observation, or a set of independent formation processes for each star system \citep{Dymont:2022, Mansfield:2022, Edwards:2023}. Therefore, a significant increase to the number of well-characterized systems is required in order to understand whether atmospheric evolution is truly stochastic or is predictable from the properties of the host system.

The recent launch of the James Webb Space Telescope (\textit{JWST}) has vastly improved the potential to characterize exoplanet systems. Studies with NIRISS/SOSS on \textit{JWST} have broken significant degeneracies typically found in ground-based observations and have allowed for the characterization of atoms and molecules such as H, He, CO, H$_2$O, and CO$_2$ in Jovian atmospheres \citep{Fu:2022, Feinstein:2022}. However, \textit{JWST} time is expensive and oversubscribed. For example, the oversubscription rate in the \textit{JWST} Cycle 3 Guest Observer call was 9 \citep{JWST:sub}. The competitive nature of the observatory creates a scenario where the use of \textit{JWST} for population studies may be unrealistic. 

Finally, the Atmospheric Remote-sensing Infrared Exoplanet Large-survey (\textit{ARIEL}) is expected to launch in 2029 with a nominal 4-year prime mission. The telescope will provide low-to-medium resolution transit spectrosocpy for a large number of exoplanets ($\sim1000$) to search for several elemental and molecular species in the atmospheres of a broad selection of exoplanet types \citep{Tinetti:2018, Eccleston:2024}. While \textit{ARIEL}'s survey has the potential to probe questions about the composition, formation, and evolution of exoplanets (as well as supplement observations from \textit{JWST}) the mission is still 5 years from launch and nearly a decade away from the end of the prime mission. Therefore, contemporary studies of exoplanet populations would benefit from rapid, low-cost, ground-based, reconnaissance observations of the exoplanets planned to be targeted with both \textit{ARIEL} and \textit{JWST}.

A new ground-based observing technique called common-path multi-band imaging (CMI) is capable of achieving self-referenced differential photometric precision comparable to space-based telescopes using modest ground-based telescopes ($\sim300$~ppm (0.03\%) at 60~s cadence on a 2~m class telescope) \citep{Schmidt:2022, Limbach:2024, Schmidt:2024}. The Exoplanet Transmission Spectroscopy Imager (ETSI) was designed to make use of the CMI technique to enable ground-based exoplanet transmission spectro-photometry from small to mid-sized observatories. ETSI has been in operation since 2022 on the McDonald Observatory 2.1~m Otto Struve telescope.

This manuscript details an evaluation of ETSI's capability to provide reconnaissance measurements of exoplanet atmospheres using the observations of 21 exoplanets during transit obtained as a part of instrument commissioning. We describe our efforts to confirm the ground-based signals are genuine and we provide a discussion on initial correlations found between each of the 21 measurements. This manuscript is organized as follows: Section~\ref{sec:observations} describes the CMI method, ETSI, and the selection of the exoplanet sample; Section~\ref{sec:data} describes the data processing pipeline and ETSI's achieved photometric precision; Section~\ref{sec:depths} describes how the transit depths are measured at each wavelength; Section~\ref{sec:analysis} describes our results, our uncertainty estimations, and our comparison with previous studies; Section~\ref{sec:discussion} describes how we prioritize each of the observed exoplanets for future study and discusses our initial investigations of correlations between atmospheric measurements; and Section~\ref{sec:summary} is a summary of our findings.

\section{Observations \label{sec:observations}}

\subsection{Common-Path Multi-Band Imaging (CMI)}
Most successful detections of features in exoplanet atmospheres have been obtained by large-aperture ground-based telescopes (8-10~m class) or space-based observatories. However, these facilities typically have competitive observing queues and expensive observing overheads \citep{Kreidberg:2015, Kreidberg:2018, Perryman:2018}. In addition, transmission spectroscopy usually requires the use of high-resolution spectrographs. These instruments tend to have larger uncertainties due to detector noise, achromatic sky background, and instrument flexure \citep{Limbach:2020}. These uncertainties can be compounded by the effects of stellar activity and atmospheric scintillation leading to irreducible errors and ambiguous atmospheric detections \citep{Espinoza:2019, Moran:2023, Saba:2024}. 

In contrast, CMI uses a combination of an interference filter and prism to generate a large number of discrete point spread functions (PSFs) for several spectro-photometric bandpasses. The CMI approach, therefore, only requires accurate aperture photometry to identify a broad atmospheric signal, rather than a full spectrum extraction, since each wavelength resolution appears as a point source on a given image and the wavelength solution of each bandpass is known \textit{a priori}. The flux of each PSF can then be measured and then ratio-ed to obtain a relative photometric color of the exoplanet during transit or eclipse. These colors are insensitive to most sources of systematic errors because the light has traveled through a common-path prior to entering the filter and prism \citep{Limbach:2020, Schmidt:2022}. This novel setup is capable of producing spectra (typically $400<\lambda<1000$~nm) with resolutions up to $R\approx60$, which is comparable to other exoplanet spectroscopy methods \citep{2016SPIE.9908E..38W}. 

\subsection{The Exoplanet Transmission Spectroscopy Imager (ETSI)}
The full instrumentation details of ETSI can be found in \citet{Schmidt:2024} but we briefly the instrument here. The ETSI instrument is best described as a spectro-photometer. The instrument uses a filter and prism in combination to split the incoming light into several bandpasses which are displayed on one of two cameras. These images are visually similar to images of prism spectroscopy. The instrument was designed with the intent to provide reconnaissance observations of exoplanet atmospheres, but it is not meant to provide robust abundance measurements of specific elements and molecules. Instead, ETSI observations are meant to inform the observer whether the features of an atmosphere may exist through the detection of the signatures of strong absorption features (e.g., Na, H$_2$O, etc.), or due to the statistically correlated shape of the transmission spectra (e.g. Rayleigh Scattering). Perhaps most importantly, ETSI can be used to determine whether the targeted planet's atmospheric spectra is flat. A flat atmospheric spectra could indicate the target needs multiple transit observations or that it may need to be de-prioritized as a target for future follow-up with larger observatories if additional reconnaissance observations continue to be flat or inconclusive. Ultimately, we hope ETSI can provide observers a way to make more informed decisions about which exoplanets to observe with more expensive follow-up in the era of \textit{JWST}.

In ETSI's current form, 8 bandpasses are transmitted to a scientific Complementary Metal-Oxide-Semiconductor camera (sCMOS) and 7 bandpasses are reflected to a second sCMOS camera. For the remainder of this manuscript these cameras will be referred to as the transmitted camera and the reflected camera respectively. The transmitted camera is a 2048$\times$2048 pixel Andor-Marana CMOS detector with a pixel size of 11~$\mu$m, a pixel scale of 0.182 \arcsec/pix, and a field of view of  6.2\arcmin$\times$ 6.2\arcmin. There were two reflected cameras used in this study. The first camera was a 3200$\times$3200 pixel Teledyne Kinetix camera with a pixel size of 6.5~$\mu$m, a pixel scale of 0.107 \arcsec/pix, and a field of view of  5.7\arcmin$\times$ 5.7\arcmin. The second reflected camera was a 2048$\times$2048 Teledyne Kuro camera with a pixel size of 11~$\mu$m, a pixel scale of 0.182 \arcsec/pix, and a field of view of 6.2\arcmin$\times$ 6.2\arcmin.

As previously mentioned, ETSI's current configuration provides concurrent spectro-photometric measurements for 15 bandpasses. The wavelength placement of the 15 ETSI bandpasses were selected to target several features common in Hot-Jupiter-like atmospheres: sodium (Na), potassium (K), titanium oxide (TiO), methane (CH$_4$), water (H$_2$O), and Rayleigh Scattering (see \citet{Limbach:2020} for more details on the selection process). The approximate center of each bandpass and the molecular feature they are expected to cover is denoted in Table~\ref{tb:bandpasses} \citep{Limbach:2020, Schmidt:2022, Schmidt:2024}. The intensity measured in each of these bandpasses does not provide an absolute measurement for a given molecule's abundance, instead, the ratio of each bandpass intensity provides a measure of the relative feature strength in a given exoplanet's atmosphere. The process of determining the relative abundance of specific molecules is convenient because the instrument is only required to produce high precision, differential photometric measurements during transit. 

ETSI was commissioned on the 2.1~m Otto Struve Telescope at McDonald Observatory over multiple observing runs during the calendar years of 2022 and 2023. The current operational costs of the telescope ($\sim\$160$/night) are many orders of magnitude less than space-based facilities. The Otto Struve telescope is now 85 years (as of 2024), and coincidentally, was also used by Gerard Kuiper in 1944 to detect methane in the atmosphere of Saturn's moon Titan \citep{Kuiper:1944}. A handful of exoplanets, standard stars, brown-dwarfs, variable stars, and extra-galactic objects were targeted during commissioning to test the capabilities of the instrument and to compare the measurements with previous studies. The observations of the non-exoplanet targets will be described in a set of future papers.

\subsection{The Exoplanet Sample}

\begin{figure}[ht]
    \includegraphics[width=\textwidth,height=\textheight,keepaspectratio]{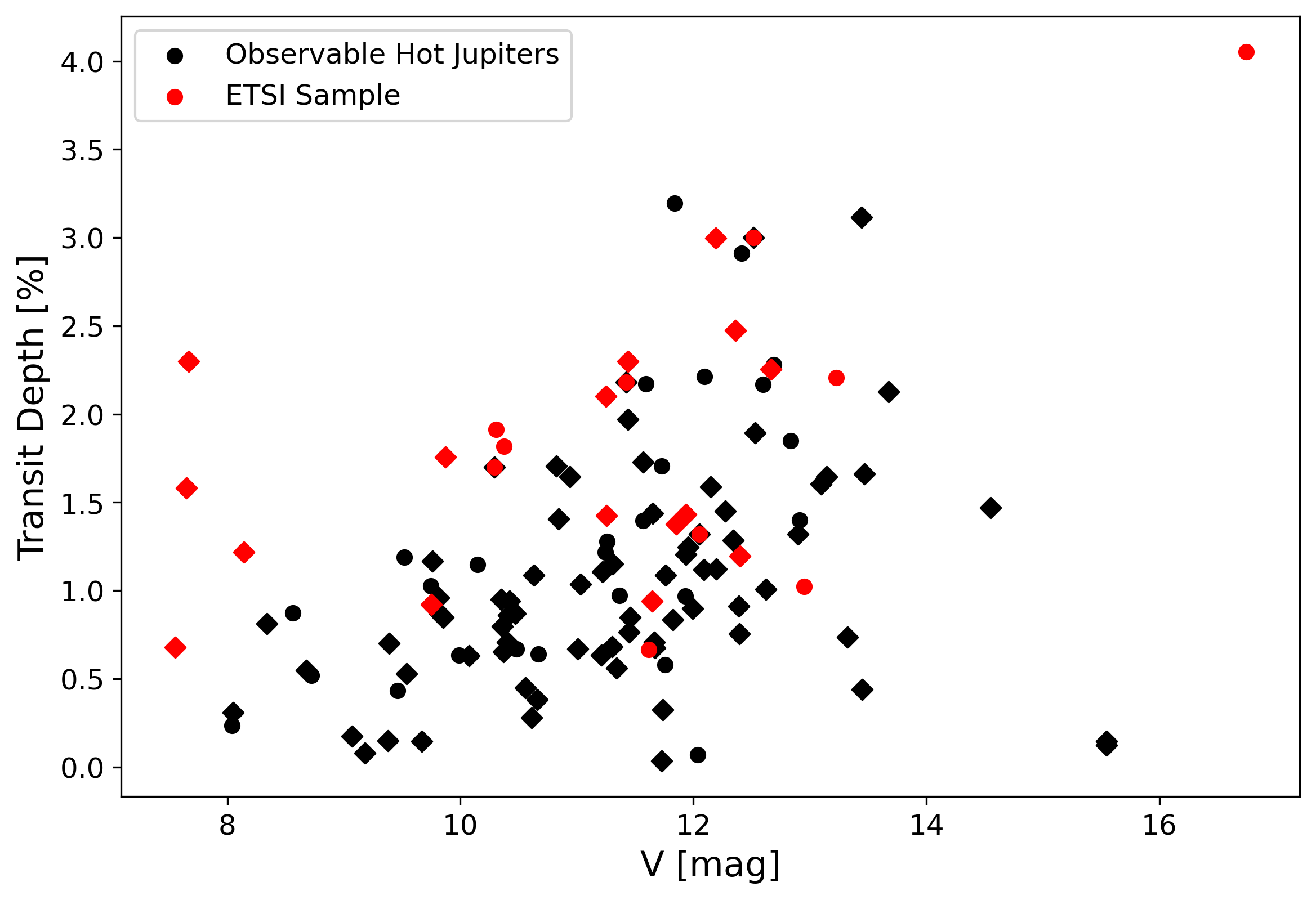}
        \caption{A comparison of the brightnesses of transiting Hot Jupiters and their transit depths for a sample of Hot Jupiters observable from McDonald Observatory ($\delta>-15^{\circ}$) with an orbital period of $P<15$~d, and a mass of $M>0.05M_J$ (black points). The sample observed as part of this work is shown with red points. Data with diamond shapes represent targets with previous transmission measurements. The majority of the observed targets in this survey have the largest transit depths for their magnitudes. Objects selected in the middle of the distribution typically have previous transmission measurements, allowing for comparison.}
        \label{fig:sample}
\end{figure}

The exoplanets analyzed as part of this study were selected to optimize our understanding of the instrument's performance during commissioning based on the following criteria. First, exoplanet targets were required to adhere to the pointing limits of the 2.1~m when ETSI was installed, typically $-30^{\circ}<\delta<60^{\circ}$\footnote{The instrument no longer has a declination limit of $\delta<60^{\circ}$ due to improvements to the instrument's housing structure which have minimized its size and decreased the 2.1~m collision limit for ETSI near the Southern and Northern Piers.}. Second, we preferred the hour angle limits of the observations to be $-3<\text{HA}<3$ to limit the effects from airmass and instrument flexure. Third, exoplanet targets were preferred to lie within a brightness range typically between $10<V<14$. ETSI has an effective brightness range of $7<V<17$, but we favored the mid-range of target brightnesses to avoid signal ambiguities which may have been caused as target brightnesses approached the instrument's saturation limit or as target brightnesses approached the level of the sky background. Fourth, exoplanet targets were preferred to have transit depths near 1\% or larger to avoid more ambiguous detections during instrument commissioning. We did violate these requirements for a subset of interesting and previously observed targets, such as HD~209458~b and KELT-9~b. Most targets were observed during the 2022 calendar year, with some additional follow-up observations occurring during the 2023 calendar year. We emphasize the main goal of this study was to stress test ETSI's capabilities. Therefore, we primarily selected targets based on their ease of observability from McDonald with preference for brighter targets with deeper transit depths. This results in a somewhat scientifically heterogeneous sample of targets. Figure~\ref{fig:sample} shows the targets selected for this study compared to the approximate sample of observable transiting Hot Jupiters which met the requirements above.

%Figure 1 - Sample Selection

In total, we attempted to observe 41 separate transits of 26 individual exoplanets during instrument commissioning. However, 8 transit observations are not included in this analysis because of poor weather conditions (i.e. heavy clouds during transit), the targets were too faint to reach a suitable signal-to-noise (i.e. Kepler-45~b), or there were minor issues with the instrument during commissioning which led to suboptimal data acquisition. This resulted in a total of 33 transit observations, which were deemed suitable for 21 individual exoplanets.

A complete list of each observed exoplanet transit with a description of the observing conditions, an approximate percentage of the observed transit length, whether the transit was observed with the reflected camera, and whether the data was used in this analysis can be found in Table~\ref{tb:exo_targets}. 

\section{Data Processing \label{sec:data}}

The ETSI data reduction pipeline automates the process of extracting exoplanet spectro-photmetry from raw images. The pipeline is written in \texttt{PYTHON} as a set of scripts, functions, and libraries that can be downloaded for inspection or use through ZENODO at \url{https://doi.org/10.5281/zenodo.14339328} and GITHUB at \url{https://github.com/ryanoelkers/etsi} \citep{etsi_pipeline}.

\subsection{Data Calibration}

The raw ETSI images were not pre-processed with bias subtraction\footnote{The bias level was removed through the ensemble background subtraction process described in Section~\ref{sec:photometry}.} or flat fielding prior to photometery. We found these calibration steps added additional noise to our photometry and because each photometric measurement is relative and the telescope was typically able to track to sub-pixel precision, these calibration steps were not needed \citep{Limbach:2020, Schmidt:2022, Limbach:2024}. 

The sCMOS detectors allow for exposures between $0.05<t<10$~s in length. While this range is useful for targeting bright ($V\sim7$) targets (because we were effectively decrease the saturation limit of the 2.1~m) these extremely short exposures vastly increased the number of images collected during a transit observation ($>100,000$ images in some cases). Therefore, most observations were co-added to one-minute timescales after data acquisition to simplify data handling. The co-added images' observation parameters (such as timing, airmass, hour angle, humidity, etc.) were calculated by taking the average of the parameters on the first and last image in a co-add sequence. The raw imagery is available upon request. 

\subsection{ETSI Photometry\label{sec:photometry}}

Light curves were extracted from ETSI images using fixed aperture photometry with elliptical apertures ($A=\pi ab$). The apertures were sized a$=40$ pixels and b$=25$ pixels for the transmitted camera and a$=65$ pixels and b$=30$ pixels for the reflected camera. The centroids for the photometry were initially placed by hand on each image because automatic finding routines were unable to reliably automatically identify the position of each bandpass due to the unique shape of the ETSI PSF. The hand-placed centroids were then automatically re-centroided using the \texttt{photutils} routine \texttt{centroids} to find the center of mass of each PSF within a box of half the size of the elliptical apertures. 

The sky background was estimated using a sky aperture of identical size to the target aperture which was placed above and below the PSF of each bandpass, typically between $150-300$~pixels of the target aperture. The sky apertures were moved closer to the target aperture or farther from the target aperture in order to avoid flux contamination from nearby stars. The fluxes in each sky aperture were summed, the two measurements were averaged, and the resulting value was subtracted from the total summed flux of the target aperture. The target flux was then converted to e$^{-}$/s using the total exposure time of the co-add (typically 60~s) and the gain of the detectors (0.61~e$^-$/ADU). The light curves were then converted to an instrumental magnitude with the standard formula

\begin{equation}
    m_{i} = 25 - 2.5\log_{10}(f)
\end{equation}

where $m_i$ is the instrumental magnitude and $f$ is the flux in e$^-$/s. 

\subsection{Removing Light Curve Systematics with Common-path Multi-band Imaging \label{sec:cmi}}

Common-path systematics (such as those from airmass, cloud cover, and atmospheric color-terms) were removed from each light curve using a time-averaged ``trend" light curve unique to each bandpass. These trend light curves were generated by linearly combining all other available spectral bandpasses (across both cameras) with the formula,

\begin{equation}
    t_{i} = \sum_{j=0}^{N} (c_j \cdot m_j + b_j) \text{; where } i \neq j
\end{equation}

where $t_i$ is the magnitude of the trend of the $i^{th}$ spectral bandpasses, $N$ is the number of bandpasses, $m_j$ is the magnitude of the $j^{th}$ bandpass, $c_j$ is the best-fit scaling correction for the $j^{th}$ bandpass, and $b_j$ is the best-fit shift for the $j^{th}$ bandpass. The final normalized light curve for each bandpass was calculated by subtracting the trend light curve from the target light curve. This type of spectral band referencing is a common method to reduce systematics in transmission light curves and is a proven technique for exoplanet light curve analysis to reduce systematic noise sources \citep{Cartier:2017, Louden:2017, Stevenson:2020, Kirk:2021, Ahrer:2022}.  

We found this method to vastly improve the capabilities of removing systematics from the target light curves over more traditional methods, such as comparison star referencing. In comparison star referencing, the light curve of a nearby bright star is subtracted from the light curve of a target star. Typically, this is a useful way to remove systematics because both stars were simultaneously observed through roughly the same atmosphere and appear at nearby positions on the detector. However, because the light from the two stars did not travel through exactly the same path prior to landing on the detector, and because not every star in the sky has a nearby, similar magnitude companion, the comparison star method is not a fool-proof method to remove systematics.

\begin{figure}[ht]
    \includegraphics[width=\textwidth,height=\textheight,keepaspectratio]{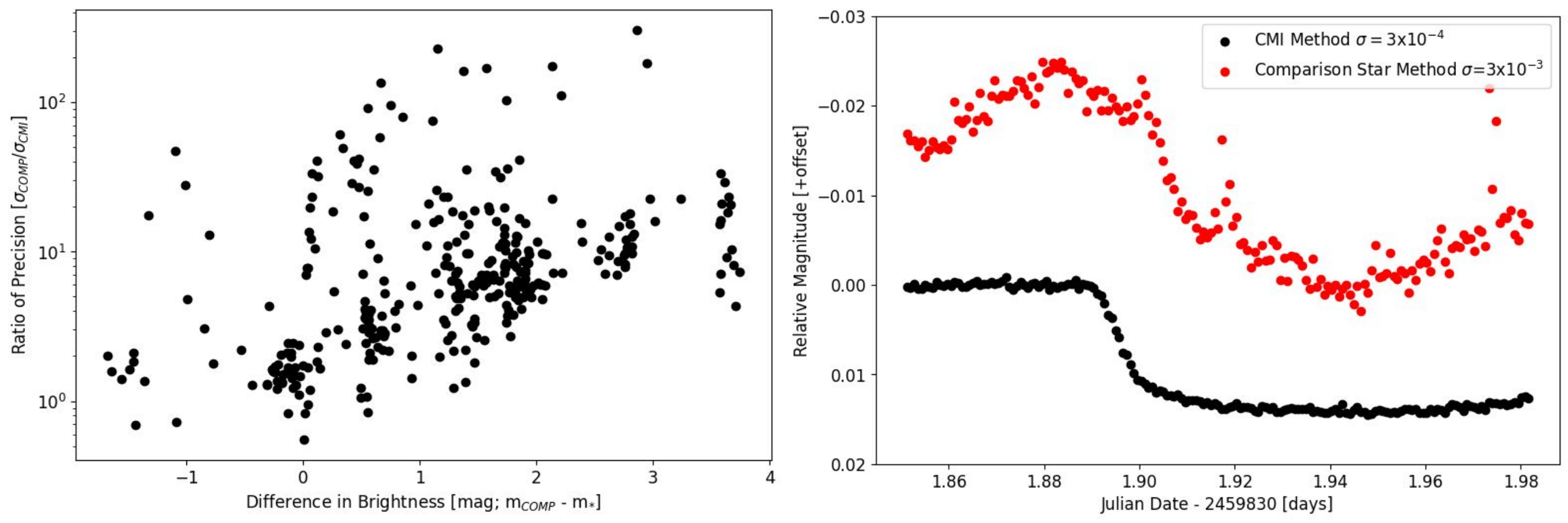}
        \caption{\textit{Left}: A comparison of the difference in brightness between the target star and comparison star and the ratio of the out-of-transit precision measured for 330 light curves observed on 22 separate nights using both the comparison star method ($\sigma_{COMP}$) and the CMI method ($\sigma_{CMI}$). On average, we find the CMI method improves light curve precision by a factor of $\sim1.5$ for comparison stars within 1 magnitude of the target star or brighter. We only found the comparison star method to improve the precision on 7 out of 330 light curves. \textit{Right}: The achieved dispersion in the light curve of WASP-33~b observed on September 08, 2022 when using the CMI method (black; $\sigma=3\times10^{-4}$) and when using a traditional comparison star method (red; $\sigma=3\times10^{-3}$). We find a factor of 10 improvement in out-of-transit precision quality when using the CMI method for this light curve.}
        \label{fig:precision}
\end{figure}

We compared the precision achieved by both the CMI method and the comparison star method for 330 light curves observed on 22 separate nights. We found, on average, the CMI method achieved better precision for ETSI light curves by a factor $\sim1.5$ when the comparison star is within 1 magnitude of the target star (or brighter) and we only found the comparison star method to provide better precision for 7 out of 330 light curves. For example, we found the dispersion in the transit light curve of WASP-33~b dropped from $\sigma=3\times10^{-3}$ to $\sigma=3\times10^{-4}$ and most of the inherent stellar variability was removed from the transit light curve when using the CMI method as compared to the comparison star method as shown in the right panel of Figure~\ref{fig:precision}. The distribution of the ratio of the achieved precision of both methods for all 330 light curves is shown in the left panel of Figure~\ref{fig:precision}. 

%Figure 2 - Precision and Transit Depths

\section{Measuring Transit Depths \label{sec:depths}}

As mentioned in Section~\ref{sec:cmi}, the common-path systematics for each light curve were removed using a trend light curve generated using all other bandpasses. While this method is effective at removing non-astrophysical signals, it also removes common, non-color-dependent astrophysical signals, such as the mean transit signal (white-light transit), because this signal is common to all bandpasses. Therefore, a model of the white-light transit signal was injected into each bandpass's de-trended light curve prior to measuring the transit depth at each wavelength. 

\begin{figure}[ht]
    \includegraphics[width=\textwidth,height=\textheight,keepaspectratio]{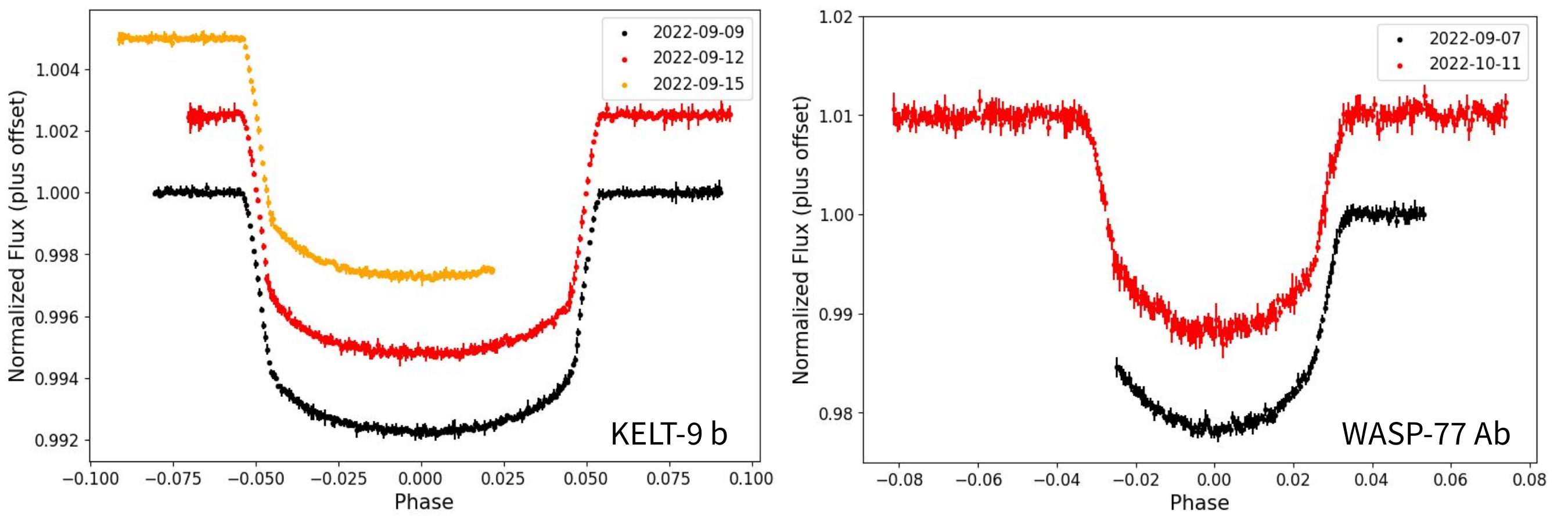}
        \caption{\textit{Left}: The white light curves of KELT-9b for observations taken across three nights. \textit{Right}: The white light curves of WASP-77~A~b for observations taken across 2 nights.}
        \label{fig:white_light}
\end{figure}

We found injecting the white-light curve transit was useful for two reasons. First, we found transit fitting routines were more capable of measuring realistic depths after a model transit was injected than they were at measuring residuals. This was particularly important for residuals containing a ``negative" transit depth because the transit was shallower than the white-light signal. Second, because we injected the same white-light transit signal into each bandpass we knew the ground truth mean transit depth \textit{a priori}. This allowed for a much more convenient comparison of the recovered signal to a flat-line, which could indicate a null-detection or possibly hazy/cloudy atmosphere. We also emphasize that this means the value of the \textit{mean} transit depth is biased towards the injected value. When executing the comparisons in Section~\ref{subsec:previous} we re-normalized the mean transit depth of the ETSI spectra to match the mean depth of previous measurements to provide a more accurate comparison of the features of each spectra. Examples of the white light curves generated for two systems are shown in Figure~\ref{fig:white_light}.

%Figure 3 - White Light Curves

The model white-light curve transit signal was generated using the \texttt{BATMAN} transit modeling software \citep{Kreidberg:2015}. The transit parameters were taken directly from NASA Exoplanet Archive \citep{NASA:transpec}\footnote{Accessed on 2023-11-15.} and the specific parameters used in this analysis are detailed in Table~\ref{tb:exo_transit}. The initial estimated quadratic limb-darkening parameters were calculated individually for each ETSI bandpass using the \texttt{LDTK} python-toolkit, the relevant stellar parameters from the literature, and the shape of the ETSI bandpasses. Finally, all light curves were converted back to flux prior-to fitting for the transit depth.

The transit depth in each bandpass was measured through a maximum likelihood estimation method where the best-fit \texttt{BATMAN} transit model was optimized over the likelihood function, 

\begin{equation}
    f(x, \bar{x})=\frac{1}{\bar{x}\cdot x \sqrt{2\pi}}\exp{(-\frac{\log^2(x)}{2\bar{x}^2})}
\end{equation}

where $x$ is the light curve data and $\bar{x}$ is the \texttt{BATMAN} model. The planet radius and the quadratic limb-darkening parameters were kept as free-parameters while all other parameters were kept constant. The planet radius to star radius ratio ($R_p/R_*$) was allowed to vary within $\pm50\%$ of the literature value and the quadratic limb-darkening parameters ($u_1$ and $u_2$) were allowed to vary between 0 and 1. Additionally, the eccentricity (\textit{e}) and the argument of periastron ($\omega$) were set to 0 since these terms were rarely provided in the literature and we wanted our methods to be consistent across all planets. Finally, if an exoplanet had observations of multiple transits, these light curves were combined into a single light curve prior to fitting for the transit depth. 

The best-fit transit depths (calculated as $(R_p/R_*)^2$) are provided in Table~\ref{tb:spectra_transmission} for the transmitted camera and in Table~\ref{tb:spectra_reflected} for the reflected camera. All transit depths are shown as a percentage. The best-fit limb-darkening parameters for the transmission camera are provided in Table~\ref{tb:spectra_transmission_u1} for $u_1$ and Table~\ref{tb:spectra_transmission_u2} for $u_2$. The best-fit limb-darkening parameters for the reflected camera are provided in Table~\ref{tb:spectra_reflected_u1} for $u_1$ and Table~\ref{tb:spectra_reflected_u2} for $u_2$.

\section{Evaluating the Efficacy of ETSI Transmission Spectroscopy Measurements\label{sec:analysis}}

\subsection{Comparisons Between Camera Measurements and Comparisons of Measurements from Separate Nights}

As previously mentioned, the majority of ETSI commissioning observations included the use of two sCMOS cameras, a transmitted camera and a reflected camera. These cameras observed roughly the same wavelength range and the bandpasses in one camera bracketed the bandpasses of the other camera as detailed in Table~\ref{tb:bandpasses}. This setup provided an opportunity to check whether each ETSI observation was internally consistent between cameras because, while specific bandpasses may show distinct absorption or emission features in one camera and not the other, the spectrum's continuum should be consistent across the full ETSI wavelength range. 

%Figure 4 - Camera Comparison
\begin{figure}[ht]
    \centering
    \includegraphics[width=\textwidth,height=\textheight,keepaspectratio]{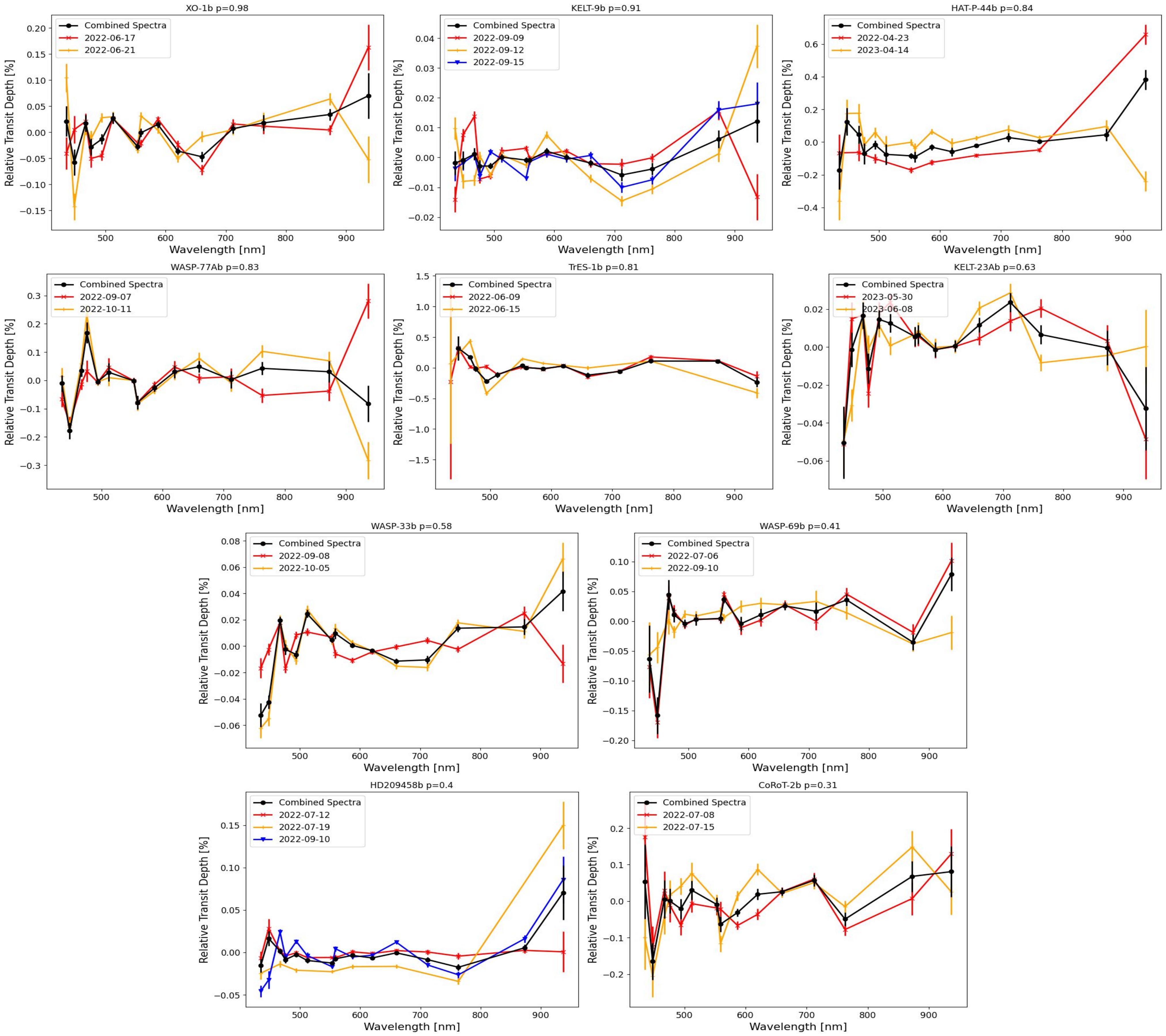}
    \caption{A visual comparison of the transmission spectra measured with ETSI on separate nights. The final combined spectra is shown as a black line and the spectra for each night are shown as colored lines (red, orange, and blue). The spectra are ordered from large p-values on the top-left to smaller p-values on the bottom-right. All exoplanets with multiple observations of their transmission spectra have large p-values ($p\geq0.31$) indicating there no exoplanets with statistically dissimilar observations even through the transmission spectra were measured days, weeks, months, and even years apart.}
    \label{fig:nightly_comp}
\end{figure}

We quantified our comparisons with a modified two-sided Anderson-Darling test \citep{Scholz:1987}. Generally speaking, the two-sided Anderson-Darling test investigates whether the null-hypothesis that the two samples are drawn from the same distribution is valid. The modified version of this test, used in the \texttt{PYTHON} library \texttt{SCIPY.STATS.ANDERSON\_KSAMP}, allows for multiple distributions to be compared without knowing the form of the underlying distribution. Anderson-Darling tests which return a p-value $<0.05$ are generally considered statistically significant enough to reject the null hypothesis and the two samples can be considered to be from separate distributions. In our analysis, tests which return $p<0.05$ suggest the two atmospheric measurements are statistically different from one another and tests with p-values where $p>0.05$ suggest we cannot statistically rule out the possibility that the two spectra are similar. All Anderson-Darling tests between the transmitted and reflected cameras never provided enough evidence to suggest the two measured spectra were statistically different from one another ($p>0.05$). 

We also separately compared the transmission spectra of exoplanets observed on multiple nights. The 10 planets observed on multiple nights included CoRoT-2~b (2 transits), HAT-P-44~b (2 transits), HD209458b (3 transits), KELT-9~b (3 transits), KELT-23~Ab (2 transits), TrES-1~b (2 transits), WASP-33b (2 transits), WASP-69~b (2 transits), WASP-77A~b (2 transits), and XO-1b (2 transits). We would expect that observations of the same exoplanet on different nights should return a similar transmission spectra on each night because the atmosphere of the planet is not expected to change over the timescale of our observations. However, discrepancies between transmission spectra measured on separate nights may be the result of detector systematics or changes in the Earth's atmosphere as these effects are likely to change from night-to-night. 

Again, each of the 10 exoplanets investigated returned a p-value of at least $p\geq0.31$ indicating the comparisons of transmission spectra measured across observational timescales of days, weeks, months, and years (in the case of HAT-P-44b) never provided enough evidence that the measured spectra were statistically different from one another. We accept these results as an indication that the transmission spectra measured with ETSI are reproducible and not significantly affected by systematics due from night-to-night variations. The transmission spectra for each of the 10 exoplanets, from each night of observation, is shown visually in Figure~\ref{fig:nightly_comp}. 

\subsection{Comparisons Between In-Transit and Out-of-Transit Observations}

While we have shown the ETSI pipeline is capable of reproducing results on a night-by-night basis and between independent cameras, there still exists the possibility that measurements taken during transit are biased by the inherent stellar variability of each exoplanet host \citep{Saba:2024}. We investigated this source of uncertainty by targeting 6 exoplanets when the planet was not in-transit. 

Our goal with these observations was to execute a bootstrap analysis by injecting the previously defined white-light transit signals (described in Section~\ref{sec:depths}) at 1000 simulated times of mid-transit during the out-of-transit observations. These simulated transit signals were subjected to the same maximum-likelihood machinery previously described in Section~\ref{sec:depths} to measure the depths in each bandpass. If the transit depths measured during this out-of-transit bootstrap simulation were inconsistent with the depths measured in-transit (through an Anderson-Darling test), then it is likely the signal measured during transit is \textit{not} due to the star's inherent color or instrument systematics. 

\begin{figure}[ht]
    \centering
    \includegraphics[width=\textwidth,height=\textheight,keepaspectratio]{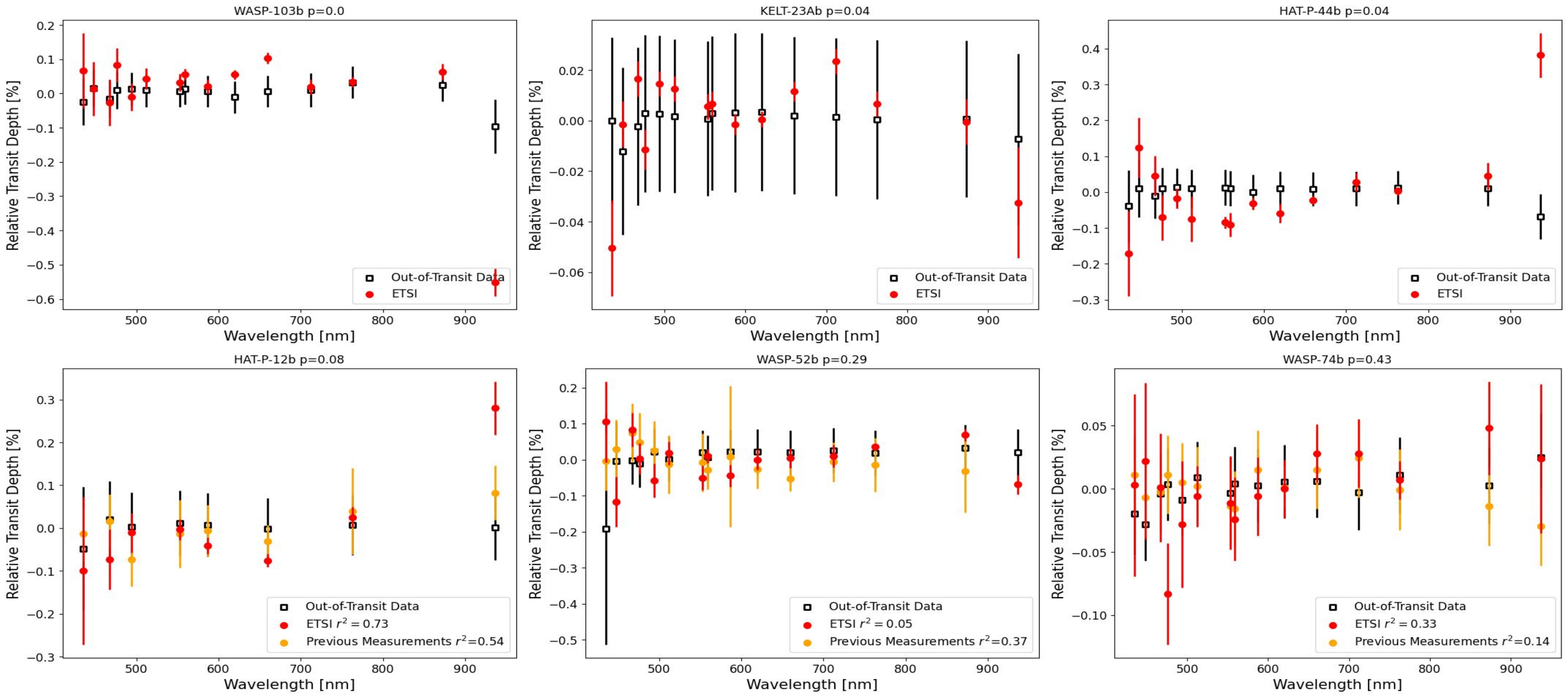}
    \caption{A visual comparison of transmission spectra measured for various planets during transit (red points) and out-of-transit (black points). The p-value of an Anderson-Darling test comparing the two measurements is shown it the title of each sub-figure. The out-of-transit measurements were generated via 1000 bootstrap simulations where the white-light transit signal (see Table~\ref{tb:exo_transit}) was injected at various times of mid-transit and measured the same way as the in-transit data. The top row of planets have a calculated p-value of $p<0.05$ indicating the in-transit and out-of-transit data are statistically different from one another. The bottom row of planets have p-values of $p>0.05$ indicating they are not statistically dis-similar to the boot-strap simulations. For these planets we additionally compared previous measurements from the Exoplanet Archive with a flat-line using an $R^2$ metric. We find the previous measurements all have an $R^2<0.54$ which suggests the atmospheric signal is likely consistent with a flat line and the consistency between in-transit and out-of-transit observations may be because the spectra is naturally featureless.}
    \label{fig:bstrap_comparisons}
\end{figure}

Figure~\ref{fig:bstrap_comparisons} shows the results of this bootstrap comparison for all 6 planets. We find 3 of the 6 planets (HAT-P-44~b, KELT-23A~b, and WASP-103~b) have $p<0.05$. This result suggests the recovered atmospheric signals are not consistent with the inherent color changes in the host star. 

%Figure 5 - Bootstrap Comparisons

HAT-P-12~b, WASP-52~b, and WASP-74~b do not pass the statistical cutoff of $p<0.05$ and have values of $p=0.08$, $p=0.29$, and $p=0.43$ respectively. Coincidentally, each of these 3 exoplanets also had previous measurements of their spectra. We used these previous measurements, in combination with a calculation of the coefficient of determination ($R^2$), to determine whether the spectrum of the atmosphere has been consistently measured to be flat. Typically, $R^2$ values closer to one indicate a linear trend with a non-zero slope better explains the data's variance than a linear trend with a slope of zero (i.e. a flat line). If the previous measurements of each planet's atmosphere return a low $R^2$ value this could be interpreted as a feature of the spectra itself and not a feature of the ETSI data reduction process \footnote{For a full investigation into comparisons between previous observations and ETSI observations the reader is directed to Section~\ref{subsec:previous}.}. 

The $R^2$ values were calculated as $R^2=0.54$, $R^2=0.37$, and $R^2=0.14$ for HAT-P-12~b, WASP-52~b, and WASP-74~b respectively. These results indicate measurements of these spectra are typically lacking strong features and the results of our Anderson-Darling test do not necessarily reflect a deficiency in ETSI's ability to detect an atmospheric signal.

\subsection{Comparisons Between ETSI, the Hubble Space Telescope, and Other Observatories\label{subsec:previous}}

We further estimated the uncertainty of the ETSI measurements by targeting 8 exoplanets which have been observed by other observatories, including the Hubble Space Telescope (\textit{HST}). While we have shown our ETSI measurements are internally consistent, if our results are biased by our methods then the efficacy of the instrument is in question because it is unable to provide an accurate estimate of the true atmospheric signal. The previous transmission spectra were drawn from the Atmospheric Spectroscopy table from the NASA Exoplanet Archive website and the sources of the measurements are shown in Table~\ref{tb:exo_comparison} \citep{NASA:transpec}\footnote{Accessed on 2023-11-15.}. While the Archive observations are not intended to be a complete selection of available previous measurements, we primarily used observations from the Archive because they are conveniently available for public use and they provide an objective selection of previous measurements for comparison. All available transmission spectra were used for the comparisons, however, we only included previous measurements if there was at least one set of measurements from a spectrograph as we found the color photometry was typically either too discrepant between previous observations or was too sparse to effectively interpolate to the ETSI bandpasses. We also do not do an ensemble comparison between previous observations if they only provided a single data point. 

We executed two comparisons with previous data. First, we compared the measured ETSI spectra with each individual previous observation through an Anderson-Darling test. The previous data was binned to the wavelengths of the ETSI measurements by taking the average of all data within the width of each ETSI bandpass. The uncertainty of each bin was defined as the standard deviation of the data points in the wavelength range. When there were less than 2 data points in a given ETSI wavelength range, the data was instead interpolated to the peak ETSI wavelength, and the uncertainty was accepted as the standard deviation of all previous data points. The results of this comparison are described below and detailed in Table~\ref{tb:exo_comparison}.

Next, we generated an ensemble comparison for each of our targets by binning all available previous data to the ETSI wavelength range. Each bin was an average of all data available for the given planet within the width of each ETSI bandpass and the uncertainty of each bin was defined as the standard deviation of the data points in the wavelength range. When there were less than 2 data points in a given ETSI wavelength range, the data was instead interpolated to the peak ETSI wavelength, and the uncertainty was accepted as the standard deviation of all previous data points. As previous mentioned, we shifted the mean transit depth of the ETSI measurements to match the mean transit depth of previous measurements prior to comparison. This was done because the mean transit depths injected into ETSI light curves could be discrepant from previous observations by $\sim0.02-0.1$\% depending on the exoplanet parameters used to model the transit. We found these discrepancies tended to bias the Anderson-Darling results due difference between the mean transit depth of the two spectra rather than the relative strengths of each feature.

%Figure 6 - ETSI vs Other Observatories 
\begin{figure}[ht]
    \centering
    \includegraphics[width=\textwidth,height=\textheight,keepaspectratio]{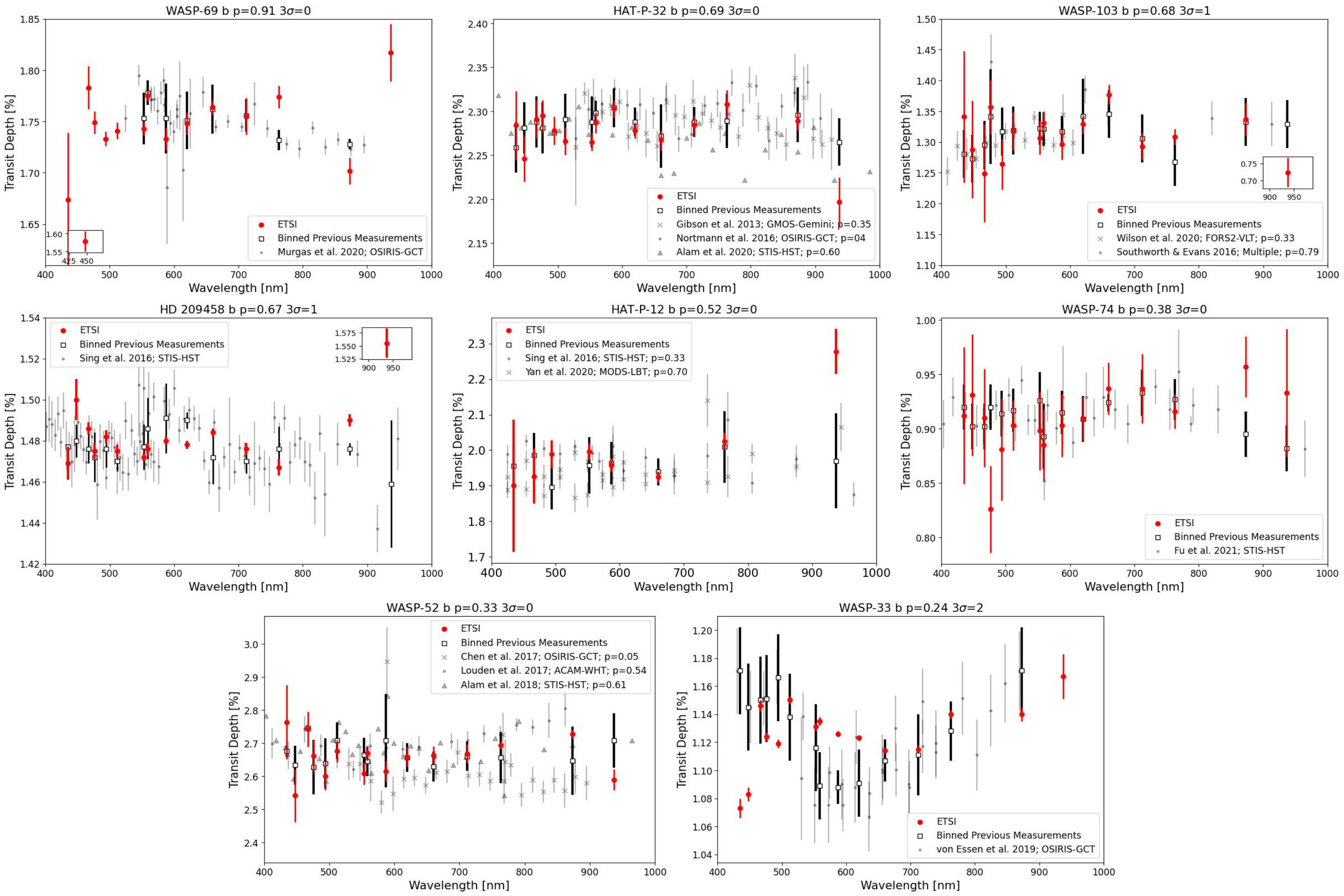}
    \caption{A visual comparison of the transmission spectra measured with ETSI (red data points) and other observatories (black points are bins and grey points are raw data). The figures are ordered from top-left to bottom-right by decreasing p-value. All comparisons between ETSI and other observatories return p-values which suggest there is not enough evidence to determine the measurements are statistically dissimilar ($p>0.05$). We also find no comparison has more than 2 points across all 15 bandpasses with a discrepancy more than 3$\sigma$.}
    \label{fig:exo_comparison}
\end{figure}

Once again, the measurements from ETSI and other observatories were statistically compared with an Anderson-Darling test \citep{Scholz:1987}. Our test showed that all 5 exoplanet spectra measured with both ETSI and HST are not statistically dissimilar ($p\geq0.33$) and all 3 exoplanet spectra measured between ETSI and other ground-based observatories are not statistically dissimilar ($p\geq0.24$). Additionally, we find no comparison has more than 2 out of 15 bandpasses which are discrepant by more than 3~sigma with 5 out of 8 comparisons having all bandpasses consistent within 3 sigma. The results for each exoplanet are discussed below and shown visually in Figure~\ref{fig:exo_comparison}. A complete list of comparison instruments, references of previous measurements, and the results of our statistical comparisons is shown in Table~\ref{tb:exo_comparison}. 

The 5 exoplanets observed in our sample with both ETSI measurements and STIS-HST measurements provided by the archive were: HD~209458~b, HAT-P-12~b, HAT-P-32~b, WASP-52~b, and WASP-74~b. The ETSI spectra of HD~209458~b is statistically consistent with observations from \citet{Sing:2016} ($p=0.67$) and it appears to show a recovery of the well-discussed sodium feature at 589~nm. Visually we find excellent agreement between both measurements for HAT-P-12~b ($p=0.52)$ with a recovered absorption feature near $\sim500$~nm in both spectra. The ETSI spectra is more statistically consistent with the LBT observations from \citet{Yan:2020} ($p=0.70$), but are well aligned with the HST observations from \citet{Sing:2016} ($p=0.33$). We find the ETSI measurements for HAT-P-32~b are in good agreement with the combined previous measurements ($p=0.62$). Visually and statistically, our results align well with the results from \citet{Gibson:2013} ($p=0.35$) and  \citet{Alam:2020} ($p=0.6$) over \citet{Mallonn:2016} ($p=0.16$) and \citet{Nortman:2016} ($p=0.04$) which show slightly deeper transits in the red end of the spectrum. The ETSI results for WASP-52~b are more visually consistent with those from \citet{Louden:2017} ($p=0.54$) and \citet{Alam:2018} ($p=0.61$), which show deeper transits redder than 770~nm, than the measurements of \citet{Chen:2017} ($p=0.05$) which show a flatter spectrum at all wavelengths and a distinct sodium feature. Finally, the \textit{STIS-HSTS} spectra of WASP-74~b is the most visually inconsistent with a deviation of the spectra at redder wavelengths and an Anderson-Darling p-value of $p=0.38$. Regardless, all 5 spectra are statistically consistent and these results indicate that ETSI is capable of reconnaissance measurements of exoplanetary atmospheres at a significantly reduced competitive cost (\$20/hr vs \$10k/hr) to available telescope time on space-based telescopes such as \textit{HST} \citep{hubble:cost}.

Similarly, 2 exoplanets (WASP-33b and WASP-69b) were observed with both ETSI and OSIRIS on the Gran Telescopio Canarias. The Anderson-Darling test for WASP-33~b returned a lower $p=0.24$. Both the \citet{vonEsson:2019} results and our measurements show deeper transits in the red and blue ends of the spectrum, however the ETSI results are much shallower in the middle of the spectrum. WASP-69~b is visually and statistically consistent with the results from \citet{Murgas:2020} ($p=0.91$), which indicate slight Rayleigh scattering in the atmosphere. 

Finally, WASP-103~b was previously measured using the FORS2 instrument on the VLT \citep{Wilson:2020} and with the DFOSC and GROND telescopes in \citet{SouthworhtEvans}. The ETSI and \citet{Wilson:2020} results are visually and statistically consistent with one another with a similar measurement of shallower transit depths at bluer wavelengths and a p-value of $p=0.33$. Similarly, the ETSI measurements at redder wavelengths show similar features to the \citet{SouthworhtEvans} measurements between $700<\lambda<900$~nm with a p-value of $p=0.79$.

We also compared our observations of KELT-9~b with previous observations, however, most of these observations were sporadic and outside of the ETSI bandpasses making it difficult to properly interpolate the spectra. We also found our KELT-9~b measurements were quite discrepant with previous results. KELT-9~b was observed with ETSI during 3 separate transits, all of which are statistically consistent with one another, but show very little change in the transit depth. It is possible the small number of previous data points across the full ETSI wavelength range is contributing to the discrepancy, however, the previous data does appear to show a significant emission feature near 656~nm which is recovered in both \citet{Cauley:2019} and \citet{Turner:2020} but not in ETSI data. Nevertheless, the KELT-9~b photometry has been cataloged and released for the community to investigate further.

\section{Discussion \label{sec:discussion}}

\subsection{Improving ETSI Measurements with More Sophisticated Photometric Data Reductions}

We note that there are additional more sophisticated, non-linear ways to remove systematics from ETSI light curves, such as Gaussian processing techniques. These methods have been shown to greatly reduce the dispersion in the light curves obtained from a variety of instruments and observatories and it is likely considered the standard way of de-trending transmission spectro-photometric data. Additionally, a more sophisticated PSF model could have been generated for the ETSI data and used to measure the flux of each bandpass in order to improve the systematic errors of the base photometry.

However, we elected to use fixed-aperture photometry and a linear-trend removal in this work in order to more robustly test the capabilities of the CMI method and the ETSI instrument without introducing more complexity to the final solutions. We initially tested various other data reduction methods, but we found these results were typically inconsistent between targets and were less reliable under poorer observing conditions. This led to a more in-homogeneous data reduction as the observations from each night had to be individually curated to produce precise results. While some of the results had improved precision over our fixed-aperture photometry method, we found no consistent, objective reason for why certain results were improved under these methods and others were not. However, we did find the fixed-aperture photometry and linear-trend removal methods consistently produced precision within our expectations regardless of observing conditions and target brightnesses. Therefore, we believe the results in this manuscript more closely reflect the robustness of the instrument and CMI method on their own, rather than the robustness of our capabilities to flexibly correct the data to an assumed understanding of what the atmospheric signals are ``supposed-to" look like.

Regardless, all of the ETSI light curves used in this work are available to the community through the \textit{Filtergraph} portal and we encourage community members to use and reduce the data as they see fit in their own efforts to extract spectra or improve photometry.

\subsection{A Prioritized List of Exoplanet Atmospheres for Observation with More-Precious Resources}

ETSI was designed to provide the astronomical community with low-cost reconnaissance observations as a way to prioritize targets of interest for further follow-up with larger observatories (such as \textit{HST}, \textit{JWST}, and eventually \textit{ARIEL}). In particular, we believe ETSI will provide a method of avoiding null or ambiguous detections with expensive equipment (see \citet{Kreidberg:2023}). 

Therefore, we have quantitatively ranked each exoplanet observed in this analysis based on their ETSI measurements through the use of an objective prioritization metric. This metric is an adaptation of the Transmission Metric (TSM) from \citet{Kempton:2018} where each exoplanet has had their host system's parameters weighted by their relative atmospheric signals. We replaced the scale factor term in the original TSM with a new term which scales the original TSM by the statistical dispersion of the transit depths measured by ETSI. Additionally, we replaced the ratio of planetary radius to stellar radius with the mean transit depth measured by ETSI. This results in a new version of the metric (TSM$_e$) using the following formula:

\begin{equation}
    TSM_e =  \overline{\delta_{Te}} \times \frac{T_{eff}}{M_p} \times 10^{-m_J/5} \times s
\end{equation}

where $\overline{\delta_{Te}}$ is the mean transit depth across all ETSI bandpasses for a given target, $T_{eff}$ is the equilibrium temperature of the planet, $M_p$ is the mass of the planet, $m_J$ is the 2MASS J band magnitude, and $s$ is the scale factor. Our new scaling factor ($s$) is defined as: 

\begin{equation}
    s = \sum_{i=0}^N (\frac{(\delta_{Te} - \overline{\delta_{Te}})^2}{\sigma_{T_e}})_i
\end{equation}

where $\delta_{Te}$ is the measured transit depth in the $i^{th}$ ETSI bandpass, $\sigma_{T_e}$ is the uncertainty of the measured transit depth in the $i^{th}$ ETSI bandpass, and $\overline{\delta_{Te}}$ is the mean transit depth across all ETSI bandpasses for a given target. For simplicity, we round the $TSM_e$ metric to the nearest integer.

We also use $s$ as a method to quantify which atmospheric spectra measured by ETSI are featureless, which may indicate ``hazy" or cloudy atmospheres during our observations. We defined spectra with $s\leq1$ as hazy/cloudy atmospheres. We find 6 planets (HAT-P-32~b, HD~209458~b, KELT-9~b, KELT-23A~b, WASP-48~b, and WASP-74~b) have $s\leq1$, indicating that the dispersion in the ETSI observations imply the exoplanet's atmosphere may have been hazy/cloudy during our observations. This means 15 planets in our sample appear to have clear (non-hazy) atmospheres. 

%Figure 7 - Priortized Spectra
\begin{figure}[ht]
    \centering
    \includegraphics[width=\textwidth,height=\textheight,keepaspectratio]{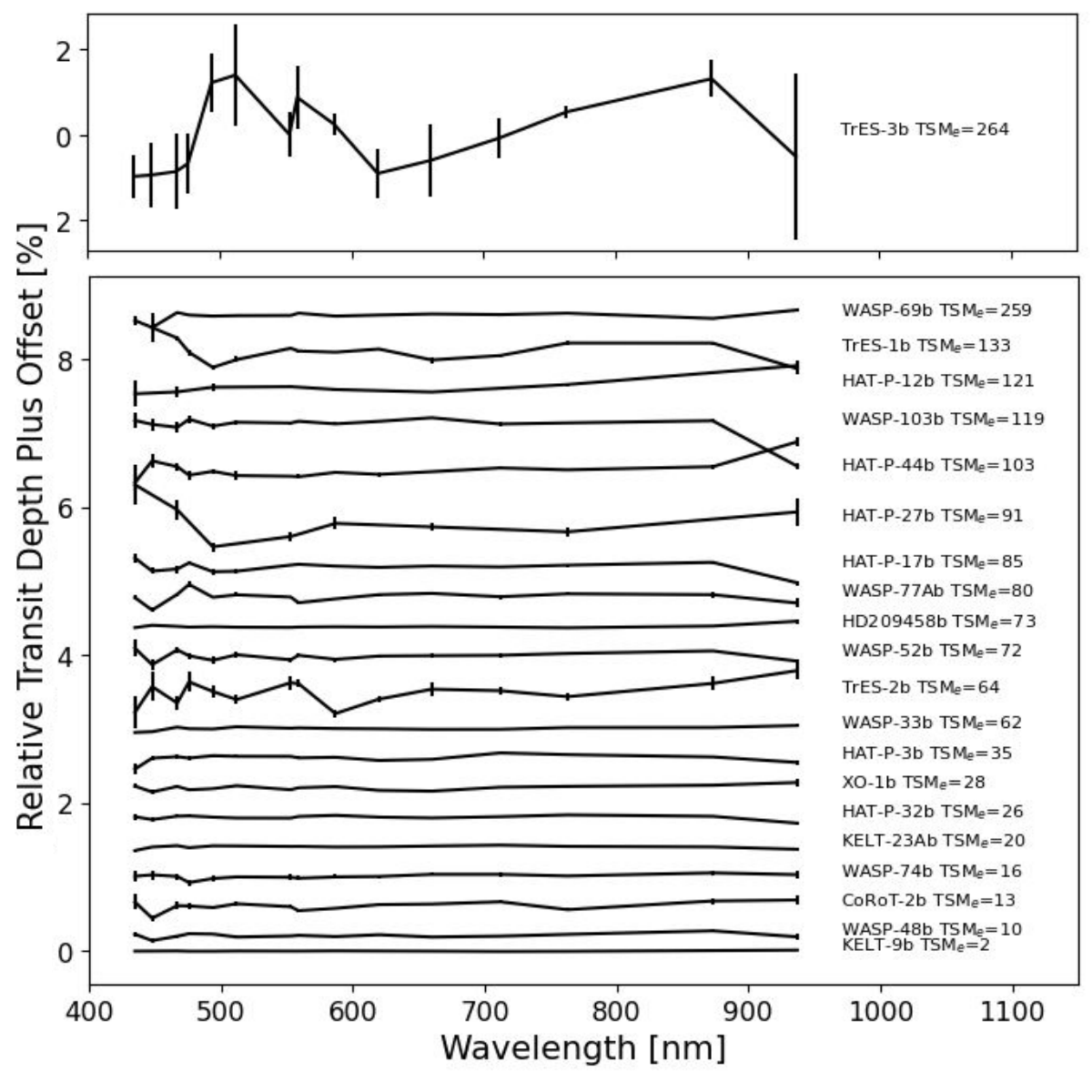}
    \caption{The transmission spectra for each of the observed exoplanets in this work, ranked by their modified transmission metric (TSM$_e$) from lowest (\textit{bottom}) to highest (\textit{top}). TrES-3~b has been plotted in a separate window to improve readability.}
    \label{fig:tsm}
\end{figure}

The calculated $TSM_e$ metric for all planets, and all parameters required for the calculations, are available in Table~\ref{tb:metric}. Additionally, Figure~\ref{fig:tsm} shows all measured ETSI spectra ranked by their respected TSM$_e$ metrics. The final spectra and their priorities are maintained as a living database on the Filtergraph visualization portal at the URL \url{https://filtergraph.com/etsi} \citep{Burger:2013}. 

\subsection{Possible Avenues for Future Work with ETSI \label{subsec:relationships}}

\begin{figure}[ht]
    \centering
    \includegraphics[width=\textwidth,height=\textheight,keepaspectratio]{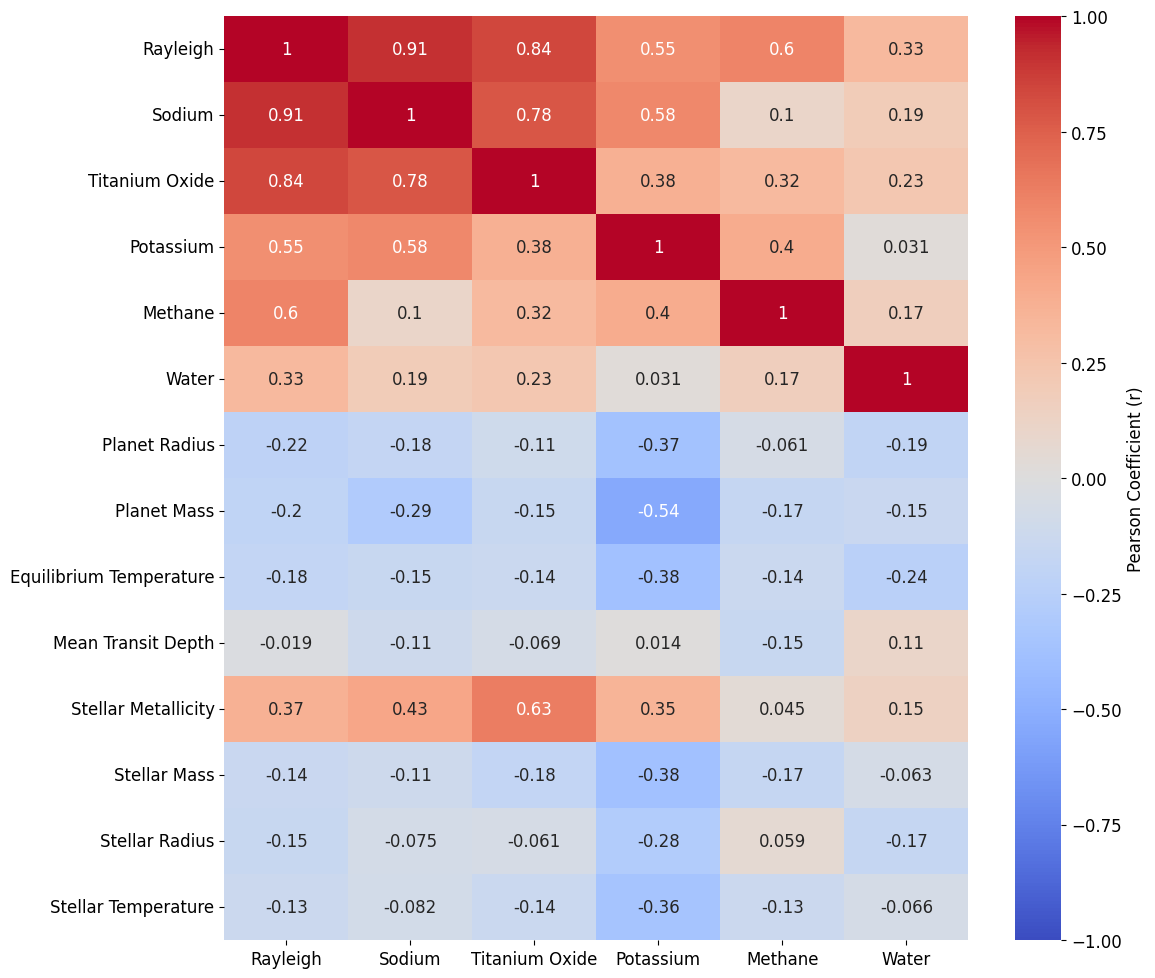}
    \caption{The Pearson (r) correlation matrix for strengths of various molecular features measured by ETSI and a set of host-system parameters. Generally, we consider any correlation coefficient of $|r|>0.5$ to be a strong relationship.}
    \label{fig:correlation matrix}
\end{figure}

We also executed an ensemble comparison of the relative strength of several molecular features to investigate whether we could detect any significant correlations between the measured spectra and the host-system parameters. First, we mean-combined the transit depths measured for each investigated molecule according to the bandpass divisions described in Table~\ref{tb:bandpasses}. Next, we normalized each molecule's transit depth by the mean transit depth of the reference bandpasses as a way to remove the bias from the spectra's continuum. Finally, we calculated the Pearson correlation coefficient ($r$) between each molecular species and between each molecular species and several host system parameters. The results of the ensemble comparison are shown in Figure~\ref{fig:correlation matrix}.

Through the course of this analysis we noticed that TrES-3~b significantly contributed to the calculated Pearson correlation coefficient. We elected to remove TrES-3~b from this analysis because we felt in some cases it was biasing our sample. For example, when TrES-3~b is included in the analysis, we calculated the Pearson correlation coefficient between the mean transit depth and the relative titanium oxide feature strength to be $r=-0.41$. However, when we removed TrES-3~b the coefficient dropped to $r=-0.069$. Conversely, the relationship between planet mass and the strength of the potassium feature was $r=-0.56$ when TrES-3~b was included and $r=-0.54$ when TrES-3~b was excluded. We interpreted this as an indication that TrES-3~b is likely an outlier in our observations and it was removed from our tests. 

%Figure 8 - Relative Intensities vs various system parameters

We find most comparisons between the strength of various molecular features and host-system parameters produce weak correlations ($|r|<0.25$). However, we do find the potassium feature is slightly anti-correlated with stellar mass ($r=-0.38$), stellar radius ($r=-0.28$), and stellar temperature ($r=-0.36$). We also find slight correlations between the stellar host metallicity and the relative strength of the sodium feature ($r=0.43$), potassium feature ($r=0.37$), and the Rayleigh scattering feature ($r=0.35$). We also find a slight anti-correlation between the potassium feature and planetary radius ($r=-0.37$), the planetary mass and the sodium feature ($r=-0.29$), and a weak anti-correlation between equilibrium temperature and the potassium feature ($r=-0.38$).

We find a stronger anti-correlation between planetary mass and the potassium feature ($r=-0.54$) and an even stronger correlation between the titanium oxide feature and the stellar host metallicity ($r=0.63$). We also note that all features (excluding water) seem to show strong positive correlations with Rayleigh scattering. Similarly, the strength of the sodium feature also correlates with titanium oxide and potassium. 

While the relationships described above may provide insight into the possible relationship between host-system environments and the atmospheric makeup of exoplanets, the sample of exoplanets studied in this manuscript (21) is too small to provide a statistical result and these correlation coefficients are only provided as a proof-of-concept for a larger study. Therefore, we calculated the sample size which should provide a statistical result using a z-test and find a sample size of 61 planets are required to achieve a 10\% margin of error and 90\% confidence interval on our results. We already have observations of 21 targets through this work and therefore would need additional observations of 40 exoplanets.

At the time of writing this manuscript there are 554 known Hot-Jupiter-like planets according to the exoplanet archive ($>0.3$~M$_{J}$, $P<10$~d) \citep{NASA:exo}\footnote{Accessed on 2024-07-23 at 11:28.} We find 108 of these 554 Hot-Jupiter-like planets orbit bright stars ($V<13.5$), are observable from McDonald Observatory ($\delta > -15^{\circ}$), have a deep transit depth ($\delta_T>0.5\%$), do not have previous transmission spectroscopy measurements, and have a full complement of infrared magnitudes (2MASS JHK and WISE W1-W4), \textit{Gaia} parallaxes, and radial velocity semi-amplitudes allowing for full system characterizations \citep{Stevens:2017, StassunGaiaPlanets:2017}. This number is well above the required threshold of 40 planets and suggests an extended project to validate these relationships could be completed from McDonald Observatory in the future.

\section{Summary \label{sec:summary}}

We have presented a set of reconnaissance observations of exoplanet atmospheres measured with the Exoplanet Transmission Spectroscopy Imager (ETSI) during commissioning. The measurements are mostly free of systematics through the use of a novel observing technique called common-path multi-band imaging (CMI), which achieves photometric color precision on-par with space-based observations (300~ppm or 0.03\%). We find all 5 exoplanet atmospheres measured with both ETSI and the Hubble Space Telescope (\textit{HST}) are not statistically dissimilar and all 3 exoplanet atmospheres measured with ETSI and other ground based observatories are not statistically dissimilar. 

Given the consistent measurements made between the transmitted and reflected cameras, the consistency of the measurements taken on separate nights, the broad differences indicated in the boot-strap analysis, and the stark similarities between the ETSI observations and multiple other observatories we believe that reconnaissance atmospheric detections can be made with ETSI. Additionally, the combination of ETSI and the 2.1~m Otto Struve telescope allows for these observations to be made for a fraction of the observational and monetary overhead previously required for these types of observations. 

Furthermore, we find 15 out of the 21 exoplanets observed show evidence for non-cloudy/non-hazy atmospheres. An ensemble analysis of all measured atmospheres suggests a relationship may exist between the strength of the titanium oxide feature and the host star's metallicity as well as a relationship between the strength of the potassium feature and the planetary mass. We estimate that additional measurements of 40 exoplanet atmospheres with ETSI could provide insight into whether these are statistically significant relationships. 

The full set of ETSI reconnaissance observations have been uploaded to the Filtergraph data visualization portal at the URL \url{https://filtergraph.com/etsi}. The portal provides access to all ETSI light curves, transmission spectra, the TSM$_e$ ranking of each atmosphere, and all available exoplanet archive data used for the various calculations described in this manuscript. The \texttt{PYTHON} code used as part of the ETSI pipeline to reduce the raw imagery is available through the GITHUB URL \url{https://github.com/ryanoelkers/etsi} and the ZENODO URL \url{https://doi.org/10.5281/zenodo.14339328}.

\acknowledgements
RJO would like to acknowledge the incredible support of the 15 undergraduates and 6 graduate student co-authors who contributed to, and in many cases led, the more than 100 nights of observation at McDonald Observatory as ETSI was commissioned and used for science. Their dedication and professionalism was key to the success of this and other ETSI projects. RJO would also like to thank Dr.~Keivan G.~Stassun for the useful discussion which improved the quality of this manuscript. Texas A\&M University thanks Charles R. ’62 and Judith G. Munnerlyn, George P. ’40 and Cynthia Woods Mitchell, and their families for support of astronomical instrumentation activities in the Department of Physics and Astronomy. This paper includes data taken at The McDonald Observatory of The University of Texas at Austin. The authors would like to thank the McDonald Observatory Telescope Allocation Committee for the generous time awarded to this project and the McDonald Observatory support staff for their support, professionalism, patience, and hospitality when commissioning ETSI. This work was funded in part by NSF-MRI grant \#1920312. This manuscript uses data from the NASA exoplanet archive. This dataset or service is made available by the NASA Exoplanet Science Institute at IPAC, which is operated by the California Institute of Technology under contract with the National Aeronautics and Space Administration. Finally, the authors would like to thank the anonymous referee for their report which greatly improved the quality of this manuscript.

\clearpage

\begin{deluxetable}{ccc}
\tabletypesize{\scriptsize}
\tablewidth{0pt}
\tablecaption{The approximate center of the ETSI bandpasses.\label{tb:bandpasses}}
\tablehead{\colhead{Wavelength [nm]} &  \colhead{Target Molecule} & \colhead{Camera}}

\startdata
937 & Water (H$_2$O) & Transmitted\\
873 & Methane (CH$_4$) & Reflected\\
763 & Potassium (K) & Transmitted\\
713 & Titanium Oxide (TiO) & Reflected\\
660 & Titanium Oxide (TiO) & Transmitted\\
620 & Titanium Oxide (TiO) & Reflected\\
587 & Sodium (Na) & Transmitted\\
559 & Sodium (Na) & Reflected\\
533 & Reference & Transmitted\\
512 & Reference & Reflected\\
494 & Rayleigh Scattering & Transmitted\\
476 & Rayleigh Scattering & Reflected\\
467 & Rayleigh Scattering & Transmitted\\
448 & Rayleigh Scattering & Reflected\\
435 & Rayleigh Scattering & Transmitted\\
\enddata

\end{deluxetable}

\clearpage

\begin{deluxetable}{ccccccc}
\tabletypesize{\scriptsize}
\tablewidth{0pt}
\tablecaption{Exoplanet Targets \label{tb:exo_targets}}
\tablehead{\colhead{Planet Name} &  \multicolumn{2}{c}{Coordinates} & \colhead{Observation Date} & \colhead{Length of Transit} & \colhead{Usable} & \colhead{Comments} \\
 & \colhead{$\alpha$ [hh:mm:ss.ss]} & \colhead{$\delta$ [dd:mm:ss]} & \colhead{YYYY-MM-DD} & \colhead{[\%]} & [Y/N] & }

\startdata
CoRoT-2~b & 19:27:06.49 & 01:23:01 & 2022-07-08 & 100\% & Y & \\
 & & & 2022-07-15 & 100\% & Y & \\
\hline
HAT-P-3~b & 13:44:22.59 & 48:01:43 & 2022-06-11 & 100\% & Y & \\
\hline
HAT-P-12~b & 13:57:33.47 & 43:29:37 & 2022-06-18 & 82\% & Y & No reflected data \\
 & & & 2023-04-28 & -- & Y & Out-of-transit data \\
 \hline
HAT-P-17~b & 21:38:08.73 & 30:29:19 & 2022-06-11 & 28\% & Y & \\
\hline
HAT-P-23~b & 20:24:29.72 & 16:45:44 & 2023-06-06 & 50\% & N & Cloudy\\
\hline
HAT-P-27~b & 14:51:04.19 & 05:56:51 & 2022-04-24 & 80\% & Y & No reflected data\\
\hline
HAT-P-32~b & 02:04:10.28 & 46:41:16 & 2022-10-04 & 50\% & Y & Partly cloudy\\
\hline
HAT-P-44~b & 14:12:34.57 & 47:00:53 & 2022-04-23 & 100\% & Y & No reflected data\\
 & & & 2023-04-10 & -- & Y & Out-of-transit data\\
 & & & 2023-04-14 & 80\% & Y & High winds\\
 \hline
HD-189733~b & 20:00:43.71 & 22:42:39 & 2022-06-12 & -- & N & No data collected\\
\hline
HD-204598~b & 22:03:10.77 & 18:53:04 & 2022-07-12 & 92\% & Y & \\
 & & & 2022-07-19 & 56\% & Y & Partly cloudy during transit\\
 & & & 2022-09-10 & 100\% & Y & Partly cloudy during transit\\
 \hline
KELT-9~b & 20:31:26.35 & 39:56:20 & 2022-07-11 & -- & N & Eclipse data; No reflected data\\
 & & & 2022-09-09 & 100\% & Y & Cloudy before transit\\
 & & & 2022-09-12 & 100\% & Y & \\
 & & & 2022-09-15 & 100\% & Y & \\
\hline
KELT-23A~b & 15:28:35.19 & 66:21:32 & 2023-05-29 & -- & Y & Out-of-transit data\\
 & & & 2023-05-30 & 100\% & Y & Cloudy after transit\\
 & & & 2023-06-08 & 100\% & Y & \\
\hline
Kepler-45~b & 19:31:29.50 & 41:03:51 & 2022-07-10 & 100\% & N & Target too faint\\
\hline
TrES-1~b & 19:04:09.85 & 36:37:57 & 2022-06-09 & 100\% & Y & \\
 & & & 2022-06-15 & 95\% & Y & No reflected data\\
 & & & 2022-09-08 & 100\% & N & Cloudy\\
\hline
TrES-2~b & 19:07:14.05 & 49:18:59 & 2022-06-07 & 100\% & Y & Partly cloudy\\
\hline
TrES-3~b & 17:52:07.02 & 37:32:46 & 2022-07-16 & 100\% & Y & \\
\hline
WASP-33~b & 02:26:51.06 & 37:33:02 & 2022-09-08 & 75\% & Y & \\
 & & & 2022-10-05 & 100\% & Y & Partly cloudy during transit\\
 \hline
WASP-48~b & 19:24:38.96 & 55:28:23 & 2022-06-10 & 100\% & Y & \\
\hline
WASP-52~b & 23:13:58.76 & 08:45:41 & 2022-07-09 & 86\% & Y &\\ 
 & & & 2022-10-13 & -- & Y & Out-of-transit data; Partly cloudy\\
 \hline
WASP-69~b & 21:00:06.20 & -05:05:40 & 2022-07-06 & 100\% & Y & \\
 & & & 2022-09-10 & 50\% & Y & \\
 \hline
WASP-74~b & 20:18:09.32 & -01:04:33 & 2022-07-18 & 100\% & Y & \\
 & & & 2023-06-07 & -- & Y & Out-of-transit data\\
 \hline
WASP-77A~b & 02:28:37.23 & -07:03:38 & 2022-09-07 & 100\% & Y & \\
 & & & 2022-10-11 & 100\% & Y & \\
 \hline
WASP-90~b & 21:02:07.68 & 07:03:23 & 2022-09-07 & 100\% & N & Cloudy during transit\\
\hline
WASP-92~b & 16:26:46.10 & 51:02:28 & 2022-06-06 & -- & N & No data during transit\\
 & & & 2023-06-01 & 100\% & N & Cloudy during transit\\
 \hline
WASP-103~b & 16:37:15.58 & 07:11:00 & 2022-06-19 & 100\% & N & Corrupted data\\
 & & & 2023-06-02 & 100\% & Y & \\
 & & & 2023-06-07 & -- & Y & Out of Transit\\
 \hline
XO-1~b & 16:02:11.85 & 28:10:10 & 2022-06-17 & 100\% & Y & \\
 & & & 2022-06-22 & -- & Y & Partly cloudy\\
\enddata

\end{deluxetable}

\clearpage

\begin{deluxetable}{ccccccccc}
\rotate
\tabletypesize{\scriptsize}
\tablewidth{0pt}
\tablecaption{White-light transit depth parameters for \texttt{BATMAN} \label{tb:exo_transit}}
\tablehead{\colhead{Planet Name} & \colhead{Time of mid-Transit} & \colhead{Period} & \colhead{$R_p/R_*$} & \colhead{$a/R_*$} & \colhead{$i$} & \multicolumn{2}{c}{References}\\
 & \colhead{BJD$_{\text{TDB}}$} & \colhead{days} & & & \colhead{deg} & \colhead{Timing} & \colhead{Transit Depth}}

\startdata
CoRoT-2b & 2457683.441580 & 1.74299705 & 0.1667 & 6.7 & 87.84 & \citet{Kokori:2022} & \citet{Alonso:2008}\\
HAT-P-3b & 2456843.022438 & 2.89973815 & 0.1063 & 10.4 & 87.1 & \citet{Kokori:2022} & \citet{Chan:2011}\\
HAT-P-12b & 2456851.481119 & 3.21305762 & 0.1406 & 11.77 & 89 & \citet{Kokori:2022} & \citet{Hartman:2009}\\
HAT-P-17b & 2456703.460703 & 10.33853522 & 0.1238 & 22.6 & 89.2 & \citet{Kokori:2022} & \citet{Howard:2012}\\
HAT-P-27b & 2457128.310660 & 3.03957804 & 0.119 & 9.7 & 84.7 & \citet{Kokori:2022} & \citet{Beky:2011}\\
HAT-P-32b & 2456265.154123 & 2.15000820 & 0.1489 & 5.34 & 89 & \citet{Kokori:2022} & \citet{Wang:2019}\\
HAT-P-44b & 2457679.786450 & 4.30119043 & 0.1343 & 11.5 & 89.1 & \citet{Kokori:2022} & \citet{Hartman:2014}\\
HD209458b & 2455420.844560 & 3.52474955 & 0.12086 & 8.76 & 86.71 & \citet{Kokori:2022} & \citet{Torres:2008}\\
KELT-9b & 2458955.970923 & 1.48111874 & 0.08228 & 3.2 & 86.79 & \citet{Kokori:2022} & \citet{Gaudi:2017}\\
KELT-23Ab & 2458918.461247 & 2.25528745 & 0.132 & 7.556 & 85.96 & \citet{Kokori:2022} & \citet{Maciejewski:2020}\\
TrES-1b & 2456822.891157 & 3.03006948 & 0.1358 & 10.52 & 90 &  \citet{Kokori:2022} & \citet{Torres:2008}\\
TrES-2b & 2454849.526640 & 2.47061892 & 0.1278 & 8.06 & 84.07 & \citet{Kipping:2011} & \citet{Kipping:2011}\\
TrES-3b & 2457585.914587 & 1.30618635 & 0.1582 & 6.731 & 81.85 &  \citet{Kokori:2022} & \citet{Saeed:2020}\\
WASP-33b & 2454163.224510 & 1.21986690 & 0.1066 & 3.788 & 87.67 & \citet{CollierCameron:2010} & \citet{CollierCameron:2010}\\
WASP-48b & 2458106.263140 & 2.14363679 & 0.0958 & 4.7 & 82 & \citet{Kokori:2022} & \citet{Ciceri:2015}\\
WASP-52b & 2456784.057988 & 1.74978117 & 0.16378 & 7.38 & 85.15 & \citet{Kokori:2022} & \citet{Mancini:2017}\\
WASP-69b & 2457269.013220 & 3.86813888 & 0.1336 & 11.953 & 86.71 & \citet{Kokori:2022} & \citet{Anderson:2014}\\
WASP-74b & 2457103.325971 & 2.13775367 & 0.098 & 4.861 & 79.81 & \citet{Kokori:2022} & \citet{Hellier:2015}\\
WASP-77Ab & 2458693.870688 & 1.36002895 & 0.1301 & 5.41 & 89.4 & \citet{Kokori:2022} & \citet{Maxted:2013}\\
WASP-103b & 2457308.324538 & 0.92554539 & 0.111 & 2.9829 & 89.22 & \citet{Kokori:2022} & \citet{Barros:2022}\\
XO-1b & 2455787.553228 & 3.94150468 & 0.1326 & 11.24 & 88.8 & \citet{Kokori:2022} & \citet{Torres:2008}\\
\enddata

\end{deluxetable}

\clearpage

\begin{deluxetable}{ccccccccc}
\tabletypesize{\scriptsize}
\tablewidth{0pt}
\tablecaption{Measured ETSI transmission spectra from the transmitted camera. All values are in percentage.\label{tb:spectra_transmission}}
\tablehead{\colhead{Planet Name} & \colhead{937 nm} & \colhead{763 nm} & \colhead{660 nm} & \colhead{587 nm} & \colhead{553 nm} & \colhead{494 nm} & \colhead{467 nm} & \colhead{435 nm}}

\startdata
CoRoT-2b & 2.861 $\pm$ 0.07 & 2.731 $\pm$ 0.018 & 2.806 $\pm$ 0.011 & 2.749 $\pm$ 0.011 & 2.771 $\pm$ 0.019 & 2.759 $\pm$ 0.028 & 2.785 $\pm$ 0.052 & 2.834 $\pm$ 0.102\\
HAT-P-3b & 1.055 $\pm$ 0.026 & 1.153 $\pm$ 0.008 & 1.095 $\pm$ 0.009 & 1.12 $\pm$ 0.01 & 1.131 $\pm$ 0.011 & 1.141 $\pm$ 0.017 & 1.127 $\pm$ 0.035 & 0.973 $\pm$ 0.075\\
HAT-P-12b & 2.267 $\pm$ 0.062 & 2.015 $\pm$ 0.024 & 1.915 $\pm$ 0.015 & 1.949 $\pm$ 0.019 & 1.987 $\pm$ 0.024 & 1.979 $\pm$ 0.046 & 1.918 $\pm$ 0.07 & 1.891 $\pm$ 0.172\\
HAT-P-17b & 1.301 $\pm$ 0.031 & 1.539 $\pm$ 0.02 & 1.527 $\pm$ 0.012 & 1.526 $\pm$ 0.009 & 1.54 $\pm$ 0.014 & 1.45 $\pm$ 0.041 & 1.487 $\pm$ 0.052 & 1.635 $\pm$ 0.059\\
HAT-P-27b & 1.637 $\pm$ 0.185 & 1.368 $\pm$ 0.064 & 1.435 $\pm$ 0.052 & 1.482 $\pm$ 0.08 & 1.304 $\pm$ 0.061 & 1.165 $\pm$ 0.062 & 1.666 $\pm$ 0.128 & 2.005 $\pm$ 0.268\\
HAT-P-32b & 2.135 $\pm$ 0.029 & 2.243 $\pm$ 0.014 & 2.203 $\pm$ 0.011 & 2.237 $\pm$ 0.011 & 2.201 $\pm$ 0.01 & 2.214 $\pm$ 0.014 & 2.226 $\pm$ 0.021 & 2.219 $\pm$ 0.039\\
HAT-P-44b & 2.205 $\pm$ 0.062 & 1.826 $\pm$ 0.012 & 1.801 $\pm$ 0.012 & 1.792 $\pm$ 0.018 & 1.739 $\pm$ 0.017 & 1.806 $\pm$ 0.028 & 1.869 $\pm$ 0.056 & 1.652 $\pm$ 0.118\\
HD209458b & 1.537 $\pm$ 0.032 & 1.45 $\pm$ 0.004 & 1.467 $\pm$ 0.002 & 1.463 $\pm$ 0.002 & 1.455 $\pm$ 0.002 & 1.465 $\pm$ 0.003 & 1.469 $\pm$ 0.003 & 1.452 $\pm$ 0.008\\
KELT-9b & 0.698 $\pm$ 0.007 & 0.672 $\pm$ 0.002 & 0.675 $\pm$ 0.001 & 0.681 $\pm$ 0.001 & 0.677 $\pm$ 0.001 & 0.674 $\pm$ 0.001 & 0.681 $\pm$ 0.002 & 0.676 $\pm$ 0.004\\
KELT-23Ab & 1.704 $\pm$ 0.022 & 1.743 $\pm$ 0.005 & 1.748 $\pm$ 0.004 & 1.735 $\pm$ 0.004 & 1.742 $\pm$ 0.005 & 1.751 $\pm$ 0.005 & 1.753 $\pm$ 0.007 & 1.686 $\pm$ 0.019\\
TrES-1b & 1.607 $\pm$ 0.08 & 1.953 $\pm$ 0.032 & 1.724 $\pm$ 0.039 & 1.829 $\pm$ 0.01 & 1.882 $\pm$ 0.011 & 1.623 $\pm$ 0.033 & 2.02 $\pm$ 0.026 & \\
TrES-2b & 1.893 $\pm$ 0.11 & 1.541 $\pm$ 0.056 & 1.641 $\pm$ 0.093 & 1.312 $\pm$ 0.052 & 1.729 $\pm$ 0.092 & 1.611 $\pm$ 0.075 & 1.455 $\pm$ 0.093 & 1.329 $\pm$ 0.223\\
TrES-3b & 2.155 $\pm$ 1.945 & 3.142 $\pm$ 0.137 & 2.069 $\pm$ 0.856 & 2.868 $\pm$ 0.237 & 2.644 $\pm$ 0.51 & 3.791 $\pm$ 0.671 & 1.822 $\pm$ 0.885 & 1.715 $\pm$ 0.493\\
WASP-33b & 1.178 $\pm$ 0.015 & 1.15 $\pm$ 0.003 & 1.124 $\pm$ 0.002 & 1.136 $\pm$ 0.002 & 1.141 $\pm$ 0.002 & 1.13 $\pm$ 0.003 & 1.156 $\pm$ 0.003 & 1.083 $\pm$ 0.009\\
WASP-48b & 0.898 $\pm$ 0.041 & 0.936 $\pm$ 0.005 & 0.894 $\pm$ 0.006 & 0.901 $\pm$ 0.008 & 0.91 $\pm$ 0.009 & 0.941 $\pm$ 0.011 & 0.907 $\pm$ 0.021 & 0.939 $\pm$ 0.034\\
WASP-52b & 2.603 $\pm$ 0.027 & 2.707 $\pm$ 0.01 & 2.677 $\pm$ 0.028 & 2.628 $\pm$ 0.031 & 2.621 $\pm$ 0.034 & 2.614 $\pm$ 0.047 & 2.755 $\pm$ 0.047 & 2.778 $\pm$ 0.112\\
WASP-69b & 1.847 $\pm$ 0.028 & 1.804 $\pm$ 0.01 & 1.793 $\pm$ 0.007 & 1.762 $\pm$ 0.012 & 1.772 $\pm$ 0.007 & 1.763 $\pm$ 0.007 & 1.813 $\pm$ 0.025 & 1.703 $\pm$ 0.056\\
WASP-74b & 0.98 $\pm$ 0.059 & 0.962 $\pm$ 0.015 & 0.984 $\pm$ 0.023 & 0.947 $\pm$ 0.031 & 0.943 $\pm$ 0.037 & 0.925 $\pm$ 0.05 & 0.956 $\pm$ 0.043 & 0.957 $\pm$ 0.072\\
WASP-77Ab & 1.606 $\pm$ 0.064 & 1.731 $\pm$ 0.023 & 1.737 $\pm$ 0.02 & 1.662 $\pm$ 0.013 & 1.687 $\pm$ 0.01 & 1.684 $\pm$ 0.011 & 1.723 $\pm$ 0.022 & 1.678 $\pm$ 0.029\\
WASP-103b & 0.678 $\pm$ 0.04 & 1.225 $\pm$ 0.014 & 1.29 $\pm$ 0.017 & 1.213 $\pm$ 0.02 & 1.224 $\pm$ 0.025 & 1.184 $\pm$ 0.041 & 1.17 $\pm$ 0.068 & 1.256 $\pm$ 0.11\\
XO-1b & 1.835 $\pm$ 0.044 & 1.783 $\pm$ 0.015 & 1.718 $\pm$ 0.01 & 1.78 $\pm$ 0.007 & 1.738 $\pm$ 0.007 & 1.752 $\pm$ 0.01 & 1.782 $\pm$ 0.016 & 1.786 $\pm$ 0.029\\
\enddata
\end{deluxetable}

\begin{deluxetable}{cccccccc}
\tabletypesize{\scriptsize}
\tablewidth{0pt}
\tablecaption{Measured ETSI transmission spectra from the reflected camera. All values are in percentage.\label{tb:spectra_reflected}}
\tablehead{\colhead{Planet Name} & \colhead{873 nm} & \colhead{712 nm} & \colhead{620 nm} & \colhead{569 nm} & \colhead{512 nm} & \colhead{476 nm} & \colhead{448 nm}}

\startdata
CoRoT-2b & 2.848 $\pm$ 0.042 & 2.837 $\pm$ 0.016 & 2.799 $\pm$ 0.015 & 2.717 $\pm$ 0.021 & 2.81 $\pm$ 0.026 & 2.781 $\pm$ 0.034 & 2.615 $\pm$ 0.05\\
HAT-P-3b & 1.125 $\pm$ 0.013 & 1.175 $\pm$ 0.009 & 1.081 $\pm$ 0.012 & 1.113 $\pm$ 0.019 & 1.132 $\pm$ 0.019 & 1.11 $\pm$ 0.022 & 1.109 $\pm$ 0.036\\
HAT-P-12b & -- & -- & -- & -- & -- & -- & --\\
HAT-P-17b & 1.577 $\pm$ 0.019 & 1.515 $\pm$ 0.012 & 1.509 $\pm$ 0.015 & 1.551 $\pm$ 0.018 & 1.457 $\pm$ 0.023 & 1.57 $\pm$ 0.024 & 1.458 $\pm$ 0.042\\
HAT-P-27b & -- & -- & -- & -- & -- & -- & --\\
HAT-P-32b & 2.225 $\pm$ 0.012 & 2.22 $\pm$ 0.01 & 2.213 $\pm$ 0.01 & 2.223 $\pm$ 0.014 & 2.202 $\pm$ 0.017 & 2.23 $\pm$ 0.02 & 2.182 $\pm$ 0.031\\
HAT-P-44b & 1.868 $\pm$ 0.037 & 1.851 $\pm$ 0.026 & 1.764 $\pm$ 0.027 & 1.732 $\pm$ 0.033 & 1.748 $\pm$ 0.063 & 1.753 $\pm$ 0.064 & 1.947 $\pm$ 0.084\\
HD209458b & 1.473 $\pm$ 0.004 & 1.459 $\pm$ 0.003 & 1.461 $\pm$ 0.002 & 1.459 $\pm$ 0.003 & 1.458 $\pm$ 0.003 & 1.459 $\pm$ 0.004 & 1.483 $\pm$ 0.009\\
KELT-9b & 0.688 $\pm$ 0.003 & 0.669 $\pm$ 0.002 & 0.679 $\pm$ 0.001 & 0.677 $\pm$ 0.001 & 0.679 $\pm$ 0.001 & 0.674 $\pm$ 0.002 & 0.677 $\pm$ 0.003\\
KELT-23Ab & 1.736 $\pm$ 0.009 & 1.76 $\pm$ 0.005 & 1.737 $\pm$ 0.003 & 1.743 $\pm$ 0.005 & 1.749 $\pm$ 0.005 & 1.725 $\pm$ 0.008 & 1.735 $\pm$ 0.009\\
TrES-1b & 1.952 $\pm$ 0.02 & 1.783 $\pm$ 0.02 & 1.871 $\pm$ 0.018 & 1.845 $\pm$ 0.025 & 1.728 $\pm$ 0.042 & 1.827 $\pm$ 0.045 & 2.162 $\pm$ 0.2\\
TrES-2b & 1.72 $\pm$ 0.088 & 1.621 $\pm$ 0.051 & 1.507 $\pm$ 0.047 & 1.717 $\pm$ 0.05 & 1.501 $\pm$ 0.058 & 1.741 $\pm$ 0.133 & 1.677 $\pm$ 0.198\\
TrES-3b & 3.878 $\pm$ 0.433 & 2.563 $\pm$ 0.456 & 1.78 $\pm$ 0.582 & 3.465 $\pm$ 0.733 & 3.96 $\pm$ 1.18 & 1.997 $\pm$ 0.687 & 1.744 $\pm$ 0.764\\
WASP-33b & 1.151 $\pm$ 0.006 & 1.125 $\pm$ 0.003 & 1.133 $\pm$ 0.001 & 1.145 $\pm$ 0.004 & 1.161 $\pm$ 0.003 & 1.134 $\pm$ 0.004 & 1.093 $\pm$ 0.005\\
WASP-48b & 0.992 $\pm$ 0.015 & 0.908 $\pm$ 0.011 & 0.931 $\pm$ 0.009 & 0.921 $\pm$ 0.01 & 0.894 $\pm$ 0.015 & 0.946 $\pm$ 0.017 & 0.838 $\pm$ 0.029\\
WASP-52b & 2.742 $\pm$ 0.017 & 2.682 $\pm$ 0.031 & 2.673 $\pm$ 0.028 & 2.683 $\pm$ 0.019 & 2.691 $\pm$ 0.032 & 2.675 $\pm$ 0.043 & 2.555 $\pm$ 0.07\\
WASP-69b & 1.731 $\pm$ 0.013 & 1.784 $\pm$ 0.015 & 1.778 $\pm$ 0.01 & 1.805 $\pm$ 0.006 & 1.77 $\pm$ 0.01 & 1.778 $\pm$ 0.012 & 1.606 $\pm$ 0.031\\
WASP-74b & 1.005 $\pm$ 0.037 & 0.984 $\pm$ 0.027 & 0.954 $\pm$ 0.023 & 0.929 $\pm$ 0.033 & 0.948 $\pm$ 0.024 & 0.867 $\pm$ 0.04 & 0.977 $\pm$ 0.062\\
WASP-77Ab & 1.719 $\pm$ 0.038 & 1.691 $\pm$ 0.031 & 1.719 $\pm$ 0.022 & 1.61 $\pm$ 0.024 & 1.717 $\pm$ 0.032 & 1.857 $\pm$ 0.037 & 1.51 $\pm$ 0.029\\
WASP-103b & 1.252 $\pm$ 0.023 & 1.211 $\pm$ 0.022 & 1.245 $\pm$ 0.013 & 1.246 $\pm$ 0.017 & 1.234 $\pm$ 0.031 & 1.271 $\pm$ 0.049 & 1.206 $\pm$ 0.079\\
XO-1b & 1.799 $\pm$ 0.011 & 1.772 $\pm$ 0.009 & 1.729 $\pm$ 0.007 & 1.764 $\pm$ 0.008 & 1.792 $\pm$ 0.011 & 1.737 $\pm$ 0.016 & 1.707 $\pm$ 0.025\\
\enddata
\end{deluxetable}

\begin{deluxetable}{ccccccccc}
\tabletypesize{\scriptsize}
\tablewidth{0pt}
\tablecaption{Measured ETSI limb-darkening parameter $u_1$ in the transmitted camera. \label{tb:spectra_transmission_u1}}
\tablehead{\colhead{Planet Name} & \colhead{937 nm} & \colhead{763 nm} & \colhead{660 nm} & \colhead{587 nm} & \colhead{553 nm} & \colhead{494 nm} & \colhead{467 nm} & \colhead{435 nm}}

\startdata
CoRoT-2b & 0.548 $\pm$ 0.068 & 0.541 $\pm$ 0.025 & 0.472 $\pm$ 0.044 & 0.498 $\pm$ 0.054 & 0.695 $\pm$ 0.061 & 0.492 $\pm$ 0.116 & 0.49 $\pm$ 0.183 & 0.299 $\pm$ 0.281\\
HAT-P-3b & 0.01 $\pm$ 0.209 & 0.702 $\pm$ 0.033 & 0.61 $\pm$ 0.114 & 0.053 $\pm$ 0.145 & 0.694 $\pm$ 0.099 & 0.7 $\pm$ 0.172 & 0.786 $\pm$ 0.157 & 0.595 $\pm$ 0.287\\
HAT-P-12b & 0.01 $\pm$ 0.028 & 0.509 $\pm$ 0.043 & 0.542 $\pm$ 0.096 & 0.686 $\pm$ 0.099 & 0.588 $\pm$ 0.141 & 0.835 $\pm$ 0.079 & 0.922 $\pm$ 0.089 & 1.0 $\pm$ 0.181\\
HAT-P-17b & 0.738 $\pm$ 0.219 & 0.522 $\pm$ 0.169 & 0.301 $\pm$ 0.151 & 0.678 $\pm$ 0.024 & 0.01 $\pm$ 0.05 & 0.721 $\pm$ 0.097 & 0.441 $\pm$ 0.239 & 0.518 $\pm$ 0.224\\
HAT-P-27b & 0.369 $\pm$ 0.161 & 0.01 $\pm$ 0.333 & 0.662 $\pm$ 0.269 & 0.665 $\pm$ 0.333 & 0.268 $\pm$ 0.28 & 0.429 $\pm$ 0.172 & 0.927 $\pm$ 0.386 & 1.0 $\pm$ 0.207\\
HAT-P-32b & 0.186 $\pm$ 0.152 & 0.219 $\pm$ 0.099 & 0.389 $\pm$ 0.068 & 0.597 $\pm$ 0.044 & 0.495 $\pm$ 0.074 & 0.679 $\pm$ 0.041 & 0.394 $\pm$ 0.139 & 0.703 $\pm$ 0.176\\
HAT-P-44b & 0.01 $\pm$ 0.0 & 0.368 $\pm$ 0.088 & 0.623 $\pm$ 0.052 & 0.667 $\pm$ 0.033 & 0.51 $\pm$ 0.1 & 0.456 $\pm$ 0.187 & 0.731 $\pm$ 0.077 & 0.947 $\pm$ 0.239\\
HD209458b & 0.057 $\pm$ 0.081 & 0.356 $\pm$ 0.046 & 0.491 $\pm$ 0.026 & 0.537 $\pm$ 0.028 & 0.49 $\pm$ 0.02 & 0.657 $\pm$ 0.028 & 0.732 $\pm$ 0.022 & 0.666 $\pm$ 0.084\\
KELT-9b & 0.01 $\pm$ 0.001 & 0.285 $\pm$ 0.025 & 0.167 $\pm$ 0.024 & 0.297 $\pm$ 0.018 & 0.258 $\pm$ 0.017 & 0.335 $\pm$ 0.018 & 0.42 $\pm$ 0.028 & 0.298 $\pm$ 0.084\\
KELT-23Ab & 0.11 $\pm$ 0.18 & 0.365 $\pm$ 0.05 & 0.485 $\pm$ 0.043 & 0.498 $\pm$ 0.047 & 0.539 $\pm$ 0.049 & 0.571 $\pm$ 0.055 & 0.735 $\pm$ 0.031 & 0.464 $\pm$ 0.195\\
TrES-1b & 0.01 $\pm$ 0.014 & 0.594 $\pm$ 0.052 & 0.161 $\pm$ 0.069 & 0.493 $\pm$ 0.056 & 0.707 $\pm$ 0.032 & 0.966 $\pm$ 0.105 & 0.546 $\pm$ 0.142 & 0.877 $\pm$ 0.416\\
TrES-2b & 0.01 $\pm$ 0.05 & 0.124 $\pm$ 0.121 & 0.01 $\pm$ 0.225 & 0.01 $\pm$ 0.06 & 0.717 $\pm$ 0.322 & 0.728 $\pm$ 0.279 & 0.476 $\pm$ 0.166 & 0.496 $\pm$ 0.312\\
TrES-3b & 0.01 $\pm$ 0.425 & 0.816 $\pm$ 0.058 & 0.01 $\pm$ 0.257 & 0.835 $\pm$ 0.094 & 0.802 $\pm$ 0.276 & 1.0 $\pm$ 0.197 & 0.212 $\pm$ 0.368 & 0.303 $\pm$ 0.296\\
WASP-33b & 0.01 $\pm$ 0.006 & 0.263 $\pm$ 0.026 & 0.355 $\pm$ 0.021 & 0.443 $\pm$ 0.021 & 0.47 $\pm$ 0.018 & 0.457 $\pm$ 0.026 & 0.626 $\pm$ 0.017 & 0.302 $\pm$ 0.074\\
WASP-48b & 0.01 $\pm$ 0.035 & 0.187 $\pm$ 0.045 & 0.01 $\pm$ 0.065 & 0.01 $\pm$ 0.062 & 0.01 $\pm$ 0.066 & 0.092 $\pm$ 0.075 & 0.01 $\pm$ 0.054 & 0.01 $\pm$ 0.011\\
WASP-52b & 0.547 $\pm$ 0.099 & 0.536 $\pm$ 0.021 & 0.472 $\pm$ 0.156 & 0.303 $\pm$ 0.175 & 0.458 $\pm$ 0.203 & 0.309 $\pm$ 0.257 & 0.928 $\pm$ 0.15 & 0.876 $\pm$ 0.261\\
WASP-69b & 0.53 $\pm$ 0.116 & 0.564 $\pm$ 0.078 & 0.698 $\pm$ 0.045 & 0.584 $\pm$ 0.082 & 0.785 $\pm$ 0.036 & 0.829 $\pm$ 0.016 & 0.916 $\pm$ 0.15 & 0.929 $\pm$ 0.22\\
WASP-74b & 0.233 $\pm$ 0.163 & 0.485 $\pm$ 0.11 & 0.622 $\pm$ 0.17 & 0.465 $\pm$ 0.214 & 0.328 $\pm$ 0.263 & 0.249 $\pm$ 0.333 & 0.756 $\pm$ 0.235 & 0.772 $\pm$ 0.371\\
WASP-77Ab & 0.077 $\pm$ 0.147 & 0.066 $\pm$ 0.104 & 0.466 $\pm$ 0.095 & 0.655 $\pm$ 0.027 & 0.58 $\pm$ 0.057 & 0.734 $\pm$ 0.028 & 0.544 $\pm$ 0.109 & 0.691 $\pm$ 0.147\\
WASP-103b & 1.0 $\pm$ 0.0 & 0.598 $\pm$ 0.041 & 0.409 $\pm$ 0.116 & 0.211 $\pm$ 0.18 & 0.563 $\pm$ 0.072 & 0.36 $\pm$ 0.221 & 0.686 $\pm$ 0.255 & 0.01 $\pm$ 0.217\\
XO-1b & 0.01 $\pm$ 0.09 & 0.452 $\pm$ 0.035 & 0.505 $\pm$ 0.06 & 0.507 $\pm$ 0.042 & 0.605 $\pm$ 0.044 & 0.739 $\pm$ 0.022 & 0.706 $\pm$ 0.071 & 0.14 $\pm$ 0.149\\
\enddata
\end{deluxetable}

\begin{deluxetable}{cccccccc}
\tabletypesize{\scriptsize}
\tablewidth{0pt}
\tablecaption{Measured ETSI limb-darkening parameter $u_1$ in the reflected camera. \label{tb:spectra_reflected_u1}}
\tablehead{\colhead{Planet Name} & \colhead{873 nm} & \colhead{712 nm} & \colhead{620 nm} & \colhead{569 nm} & \colhead{512 nm} & \colhead{476 nm} & \colhead{448 nm}}

\startdata
CoRoT-2b & 0.377 $\pm$ 0.144 & 0.454 $\pm$ 0.08 & 0.436 $\pm$ 0.069 & 0.686 $\pm$ 0.027 & 0.681 $\pm$ 0.032 & 0.202 $\pm$ 0.137 & 0.332 $\pm$ 0.201\\
HAT-P-3b & 0.03 $\pm$ 0.137 & 0.589 $\pm$ 0.045 & 0.346 $\pm$ 0.178 & 0.199 $\pm$ 0.205 & 0.808 $\pm$ 0.055 & 0.801 $\pm$ 0.223 & 0.989 $\pm$ 0.212\\
HAT-P-12b & -- & -- & -- & -- & -- & -- & -- \\
HAT-P-17b & 0.303 $\pm$ 0.13 & 0.599 $\pm$ 0.095 & 0.588 $\pm$ 0.146 & 0.57 $\pm$ 0.151 & 0.214 $\pm$ 0.158 & 0.637 $\pm$ 0.053 & 0.404 $\pm$ 0.097\\
HAT-P-27b & -- & -- & -- & -- & -- & -- & -- \\
HAT-P-32b & 0.34 $\pm$ 0.08 & 0.389 $\pm$ 0.075 & 0.474 $\pm$ 0.063 & 0.596 $\pm$ 0.089 & 0.407 $\pm$ 0.104 & 0.653 $\pm$ 0.049 & 0.288 $\pm$ 0.167\\
HAT-P-44b & 0.503 $\pm$ 0.1 & 0.51 $\pm$ 0.059 & 0.413 $\pm$ 0.16 & 0.726 $\pm$ 0.132 & 0.724 $\pm$ 0.175 & 0.8 $\pm$ 0.123 & 0.702 $\pm$ 0.197\\
HD209458b & 0.409 $\pm$ 0.03 & 0.367 $\pm$ 0.043 & 0.477 $\pm$ 0.029 & 0.55 $\pm$ 0.029 & 0.611 $\pm$ 0.041 & 0.603 $\pm$ 0.049 & 0.707 $\pm$ 0.062\\
KELT-9b & 0.131 $\pm$ 0.05 & 0.284 $\pm$ 0.035 & 0.19 $\pm$ 0.02 & 0.344 $\pm$ 0.019 & 0.316 $\pm$ 0.022 & 0.319 $\pm$ 0.028 & 0.302 $\pm$ 0.046\\
KELT-23Ab & 0.293 $\pm$ 0.089 & 0.449 $\pm$ 0.049 & 0.422 $\pm$ 0.036 & 0.552 $\pm$ 0.045 & 0.652 $\pm$ 0.042 & 0.629 $\pm$ 0.074 & 0.792 $\pm$ 0.053\\
TrES-1b & 0.01 $\pm$ 0.089 & 0.669 $\pm$ 0.027 & 0.033 $\pm$ 0.062 & 0.643 $\pm$ 0.071 & 0.858 $\pm$ 0.09 & 0.844 $\pm$ 0.12 & 0.525 $\pm$ 0.321\\
TrES-2b & 0.335 $\pm$ 0.147 & 0.01 $\pm$ 0.193 & 0.01 $\pm$ 0.219 & 0.743 $\pm$ 0.134 & 0.549 $\pm$ 0.216 & 0.775 $\pm$ 0.362 & 0.063 $\pm$ 0.31\\
TrES-3b & 0.938 $\pm$ 0.16 & 0.624 $\pm$ 0.254 & 0.01 $\pm$ 0.246 & 0.907 $\pm$ 0.322 & 0.489 $\pm$ 0.332 & 0.313 $\pm$ 0.367 & 0.01 $\pm$ 0.276\\
WASP-33b & 0.237 $\pm$ 0.049 & 0.45 $\pm$ 0.023 & 0.345 $\pm$ 0.02 & 0.513 $\pm$ 0.026 & 0.53 $\pm$ 0.023 & 0.372 $\pm$ 0.038 & 0.621 $\pm$ 0.05\\
WASP-48b & 0.01 $\pm$ 0.023 & 0.01 $\pm$ 0.116 & 0.01 $\pm$ 0.0 & 0.252 $\pm$ 0.079 & 0.263 $\pm$ 0.085 & 0.116 $\pm$ 0.076 & 0.01 $\pm$ 0.084\\
WASP-52b & 0.547 $\pm$ 0.041 & 0.497 $\pm$ 0.183 & 0.561 $\pm$ 0.167 & 0.774 $\pm$ 0.094 & 0.77 $\pm$ 0.2 & 0.871 $\pm$ 0.164 & 0.614 $\pm$ 0.321\\
WASP-69b & 0.06 $\pm$ 0.115 & 0.527 $\pm$ 0.11 & 0.637 $\pm$ 0.076 & 0.798 $\pm$ 0.013 & 0.798 $\pm$ 0.047 & 0.897 $\pm$ 0.059 & 0.158 $\pm$ 0.208\\
WASP-74b & 0.41 $\pm$ 0.241 & 0.606 $\pm$ 0.194 & 0.574 $\pm$ 0.156 & 0.337 $\pm$ 0.244 & 0.695 $\pm$ 0.145 & 0.01 $\pm$ 0.251 & 0.822 $\pm$ 0.111\\
WASP-77Ab & 0.233 $\pm$ 0.074 & 0.071 $\pm$ 0.134 & 0.638 $\pm$ 0.053 & 0.299 $\pm$ 0.13 & 0.714 $\pm$ 0.058 & 0.138 $\pm$ 0.149 & 1.0 $\pm$ 0.0\\
WASP-103b & 0.028 $\pm$ 0.126 & 0.423 $\pm$ 0.155 & 0.543 $\pm$ 0.086 & 0.449 $\pm$ 0.136 & 0.142 $\pm$ 0.16 & 0.86 $\pm$ 0.211 & 0.865 $\pm$ 0.204\\
XO-1b & 0.397 $\pm$ 0.052 & 0.306 $\pm$ 0.069 & 0.523 $\pm$ 0.053 & 0.61 $\pm$ 0.036 & 0.642 $\pm$ 0.018 & 0.555 $\pm$ 0.088 & 0.892 $\pm$ 0.082\\
\enddata
\end{deluxetable}

\begin{deluxetable}{ccccccccc}
\tabletypesize{\scriptsize}
\tablewidth{0pt}
\tablecaption{Measured ETSI limb-darkening parameter $u_2$ in the transmitted camera. \label{tb:spectra_transmission_u2}}
\tablehead{\colhead{Planet Name} & \colhead{937 nm} & \colhead{763 nm} & \colhead{660 nm} & \colhead{587 nm} & \colhead{553 nm} & \colhead{494 nm} & \colhead{467 nm} & \colhead{435 nm}}

\startdata
CoRoT-2b & 0.01 $\pm$ 0.0 & 0.01 $\pm$ 0.029 & 0.149 $\pm$ 0.076 & 0.238 $\pm$ 0.087 & 0.01 $\pm$ 0.094 & 0.352 $\pm$ 0.19 & 0.443 $\pm$ 0.303 & 0.766 $\pm$ 0.424\\
HAT-P-3b & 0.662 $\pm$ 0.298 & 0.01 $\pm$ 0.031 & 0.166 $\pm$ 0.151 & 0.852 $\pm$ 0.191 & 0.01 $\pm$ 0.133 & 0.01 $\pm$ 0.241 & 0.01 $\pm$ 0.187 & 0.847 $\pm$ 0.392\\
HAT-P-12b & 0.01 $\pm$ 0.086 & 0.01 $\pm$ 0.033 & 0.335 $\pm$ 0.158 & 0.135 $\pm$ 0.154 & 0.278 $\pm$ 0.232 & 0.01 $\pm$ 0.116 & 0.01 $\pm$ 0.081 & 0.01 $\pm$ 0.247\\
HAT-P-17b & 0.266 $\pm$ 0.321 & 0.153 $\pm$ 0.213 & 0.492 $\pm$ 0.202 & 0.01 $\pm$ 0.0 & 0.916 $\pm$ 0.092 & 0.01 $\pm$ 0.0 & 0.88 $\pm$ 0.405 & 0.01 $\pm$ 0.367\\
HAT-P-27b & 1.0 $\pm$ 0.063 & 0.795 $\pm$ 0.368 & 0.01 $\pm$ 0.293 & 0.01 $\pm$ 0.34 & 0.337 $\pm$ 0.287 & 0.01 $\pm$ 0.0 & 0.037 $\pm$ 0.44 & 0.293 $\pm$ 0.316\\
HAT-P-32b & 0.482 $\pm$ 0.233 & 0.393 $\pm$ 0.149 & 0.176 $\pm$ 0.104 & 0.019 $\pm$ 0.076 & 0.227 $\pm$ 0.112 & 0.01 $\pm$ 0.069 & 0.483 $\pm$ 0.215 & 0.32 $\pm$ 0.259\\
HAT-P-44b & 0.01 $\pm$ 0.0 & 0.274 $\pm$ 0.128 & 0.052 $\pm$ 0.082 & 0.01 $\pm$ 0.028 & 0.437 $\pm$ 0.167 & 0.426 $\pm$ 0.297 & 0.01 $\pm$ 0.032 & 0.333 $\pm$ 0.339\\
HD209458b & 0.023 $\pm$ 0.068 & 0.241 $\pm$ 0.064 & 0.078 $\pm$ 0.035 & 0.119 $\pm$ 0.037 & 0.258 $\pm$ 0.029 & 0.051 $\pm$ 0.038 & 0.023 $\pm$ 0.03 & 0.165 $\pm$ 0.119\\
KELT-9b & 0.056 $\pm$ 0.058 & 0.046 $\pm$ 0.037 & 0.266 $\pm$ 0.038 & 0.156 $\pm$ 0.028 & 0.272 $\pm$ 0.026 & 0.21 $\pm$ 0.027 & 0.13 $\pm$ 0.042 & 0.27 $\pm$ 0.137\\
KELT-23Ab & 0.557 $\pm$ 0.259 & 0.251 $\pm$ 0.069 & 0.088 $\pm$ 0.058 & 0.172 $\pm$ 0.065 & 0.199 $\pm$ 0.066 & 0.191 $\pm$ 0.074 & 0.01 $\pm$ 0.038 & 0.514 $\pm$ 0.28\\
TrES-1b & 1.0 $\pm$ 0.129 & 0.01 $\pm$ 0.0 & 1.0 $\pm$ 0.001 & 0.26 $\pm$ 0.093 & 0.014 $\pm$ 0.054 & 0.01 $\pm$ 0.184 & 0.187 $\pm$ 0.23 & 0.01 $\pm$ 0.0\\
TrES-2b & 0.874 $\pm$ 0.189 & 0.01 $\pm$ 0.046 & 1.0 $\pm$ 0.249 & 0.01 $\pm$ 0.0 & 0.01 $\pm$ 0.328 & 0.03 $\pm$ 0.282 & 0.01 $\pm$ 0.079 & 0.01 $\pm$ 0.253\\
TrES-3b & 0.01 $\pm$ 0.41 & 0.01 $\pm$ 0.001 & 0.118 $\pm$ 0.438 & 0.01 $\pm$ 0.014 & 0.038 $\pm$ 0.193 & 0.263 $\pm$ 0.2 & 0.01 $\pm$ 0.277 & 0.01 $\pm$ 0.195\\
WASP-33b & 0.405 $\pm$ 0.071 & 0.171 $\pm$ 0.041 & 0.223 $\pm$ 0.034 & 0.16 $\pm$ 0.034 & 0.174 $\pm$ 0.033 & 0.267 $\pm$ 0.042 & 0.01 $\pm$ 0.024 & 0.762 $\pm$ 0.119\\
WASP-48b & 0.01 $\pm$ 0.02 & 0.01 $\pm$ 0.01 & 0.21 $\pm$ 0.08 & 0.273 $\pm$ 0.086 & 0.209 $\pm$ 0.09 & 0.094 $\pm$ 0.103 & 0.036 $\pm$ 0.097 & 0.01 $\pm$ 0.001\\
WASP-52b & 0.01 $\pm$ 0.126 & 0.01 $\pm$ 0.0 & 0.219 $\pm$ 0.212 & 0.534 $\pm$ 0.247 & 0.44 $\pm$ 0.264 & 0.693 $\pm$ 0.335 & 0.01 $\pm$ 0.198 & 0.01 $\pm$ 0.33\\
WASP-69b & 0.01 $\pm$ 0.108 & 0.102 $\pm$ 0.096 & 0.01 $\pm$ 0.057 & 0.219 $\pm$ 0.101 & 0.01 $\pm$ 0.042 & 0.01 $\pm$ 0.018 & 0.01 $\pm$ 0.193 & 0.01 $\pm$ 0.258\\
WASP-74b & 0.01 $\pm$ 0.034 & 0.01 $\pm$ 0.105 & 0.01 $\pm$ 0.173 & 0.224 $\pm$ 0.213 & 0.473 $\pm$ 0.284 & 0.547 $\pm$ 0.344 & 0.01 $\pm$ 0.254 & 0.01 $\pm$ 0.387\\
WASP-77Ab & 1.0 $\pm$ 0.256 & 0.582 $\pm$ 0.188 & 0.213 $\pm$ 0.188 & 0.01 $\pm$ 0.026 & 0.245 $\pm$ 0.105 & 0.01 $\pm$ 0.044 & 0.333 $\pm$ 0.208 & 0.224 $\pm$ 0.245\\
WASP-103b & 1.0 $\pm$ 0.045 & 0.01 $\pm$ 0.041 & 0.189 $\pm$ 0.194 & 0.615 $\pm$ 0.283 & 0.01 $\pm$ 0.078 & 0.559 $\pm$ 0.378 & 0.01 $\pm$ 0.378 & 0.929 $\pm$ 0.379\\
XO-1b & 0.584 $\pm$ 0.176 & 0.01 $\pm$ 0.04 & 0.21 $\pm$ 0.094 & 0.146 $\pm$ 0.067 & 0.155 $\pm$ 0.072 & 0.01 $\pm$ 0.027 & 0.034 $\pm$ 0.119 & 0.939 $\pm$ 0.239\\
\enddata
\end{deluxetable}

\begin{deluxetable}{cccccccc}
\tabletypesize{\scriptsize}
\tablewidth{0pt}
\tablecaption{Measured ETSI limb-darkening parameter $u_2$ in the reflected camera. \label{tb:spectra_reflected_u2}}
\tablehead{\colhead{Planet Name} & \colhead{873 nm} & \colhead{712 nm} & \colhead{620 nm} & \colhead{569 nm} & \colhead{512 nm} & \colhead{476 nm} & \colhead{448 nm}}

\startdata
CoRoT-2b & 0.148 $\pm$ 0.239 & 0.117 $\pm$ 0.126 & 0.201 $\pm$ 0.114 & 0.01 $\pm$ 0.036 & 0.01 $\pm$ 0.028 & 0.839 $\pm$ 0.208 & 0.884 $\pm$ 0.313\\
HAT-P-3b & 0.661 $\pm$ 0.19 & 0.01 $\pm$ 0.049 & 0.365 $\pm$ 0.255 & 0.682 $\pm$ 0.289 & 0.01 $\pm$ 0.0 & 0.034 $\pm$ 0.289 & 0.01 $\pm$ 0.309\\
HAT-P-12b & -- & -- & -- & -- & -- & -- & --\\
HAT-P-17b & 0.111 $\pm$ 0.178 & 0.03 $\pm$ 0.128 & 0.19 $\pm$ 0.194 & 0.181 $\pm$ 0.208 & 1.0 $\pm$ 0.214 & 0.01 $\pm$ 0.0 & 1.0 $\pm$ 0.053\\
HAT-P-27b & -- & -- & -- & -- & -- & -- & --\\
HAT-P-32b & 0.133 $\pm$ 0.118 & 0.217 $\pm$ 0.112 & 0.148 $\pm$ 0.1 & 0.05 $\pm$ 0.127 & 0.361 $\pm$ 0.164 & 0.01 $\pm$ 0.056 & 0.759 $\pm$ 0.242\\
HAT-P-44b & 0.01 $\pm$ 0.139 & 0.01 $\pm$ 0.059 & 0.481 $\pm$ 0.269 & 0.143 $\pm$ 0.206 & 0.248 $\pm$ 0.321 & 0.01 $\pm$ 0.13 & 0.01 $\pm$ 0.228\\
HD209458b & 0.01 $\pm$ 0.04 & 0.225 $\pm$ 0.058 & 0.144 $\pm$ 0.04 & 0.135 $\pm$ 0.04 & 0.152 $\pm$ 0.056 & 0.203 $\pm$ 0.069 & 0.01 $\pm$ 0.083\\
KELT-9b & 0.158 $\pm$ 0.076 & 0.104 $\pm$ 0.055 & 0.277 $\pm$ 0.03 & 0.124 $\pm$ 0.029 & 0.212 $\pm$ 0.034 & 0.258 $\pm$ 0.042 & 0.384 $\pm$ 0.07\\
KELT-23Ab & 0.232 $\pm$ 0.122 & 0.093 $\pm$ 0.066 & 0.237 $\pm$ 0.05 & 0.133 $\pm$ 0.063 & 0.068 $\pm$ 0.057 & 0.197 $\pm$ 0.104 & 0.01 $\pm$ 0.072\\
TrES-1b & 0.466 $\pm$ 0.16 & 0.01 $\pm$ 0.0 & 1.0 $\pm$ 0.085 & 0.01 $\pm$ 0.093 & 0.01 $\pm$ 0.086 & 0.01 $\pm$ 0.175 & 0.01 $\pm$ 0.33\\
TrES-2b & 0.01 $\pm$ 0.042 & 0.699 $\pm$ 0.216 & 0.61 $\pm$ 0.231 & 0.01 $\pm$ 0.134 & 0.01 $\pm$ 0.218 & 0.074 $\pm$ 0.364 & 1.0 $\pm$ 0.314\\
TrES-3b & 0.013 $\pm$ 0.096 & 0.01 $\pm$ 0.2 & 0.01 $\pm$ 0.26 & 0.225 $\pm$ 0.229 & 1.0 $\pm$ 0.437 & 0.01 $\pm$ 0.202 & 0.01 $\pm$ 0.237\\
WASP-33b & 0.124 $\pm$ 0.08 & 0.025 $\pm$ 0.035 & 0.283 $\pm$ 0.032 & 0.074 $\pm$ 0.044 & 0.063 $\pm$ 0.038 & 0.445 $\pm$ 0.057 & 0.214 $\pm$ 0.085\\
WASP-48b & 0.032 $\pm$ 0.075 & 0.312 $\pm$ 0.136 & 0.021 $\pm$ 0.052 & 0.01 $\pm$ 0.06 & 0.01 $\pm$ 0.017 & 0.01 $\pm$ 0.071 & 0.49 $\pm$ 0.192\\
WASP-52b & 0.01 $\pm$ 0.026 & 0.117 $\pm$ 0.238 & 0.13 $\pm$ 0.222 & 0.01 $\pm$ 0.125 & 0.01 $\pm$ 0.263 & 0.01 $\pm$ 0.228 & 0.475 $\pm$ 0.428\\
WASP-69b & 0.594 $\pm$ 0.139 & 0.154 $\pm$ 0.134 & 0.084 $\pm$ 0.096 & 0.01 $\pm$ 0.006 & 0.01 $\pm$ 0.055 & 0.01 $\pm$ 0.072 & 1.0 $\pm$ 0.262\\
WASP-74b & 0.165 $\pm$ 0.234 & 0.01 $\pm$ 0.196 & 0.01 $\pm$ 0.155 & 0.354 $\pm$ 0.249 & 0.01 $\pm$ 0.149 & 0.791 $\pm$ 0.261 & 0.01 $\pm$ 0.102\\
WASP-77Ab & 0.01 $\pm$ 0.0 & 0.867 $\pm$ 0.229 & 0.01 $\pm$ 0.089 & 0.755 $\pm$ 0.229 & 0.01 $\pm$ 0.045 & 0.538 $\pm$ 0.268 & 0.39 $\pm$ 0.064\\
WASP-103b & 0.755 $\pm$ 0.219 & 0.313 $\pm$ 0.267 & 0.01 $\pm$ 0.124 & 0.147 $\pm$ 0.201 & 0.869 $\pm$ 0.233 & 0.01 $\pm$ 0.308 & 0.01 $\pm$ 0.317\\
XO-1b & 0.01 $\pm$ 0.075 & 0.361 $\pm$ 0.112 & 0.197 $\pm$ 0.085 & 0.061 $\pm$ 0.06 & 0.01 $\pm$ 0.02 & 0.279 $\pm$ 0.151 & 0.01 $\pm$ 0.114\\
\enddata
\end{deluxetable}

\clearpage 

\begin{deluxetable}{cccccc}
\tabletypesize{\scriptsize}
\tablewidth{0pt}
\tablecaption{Comparisons between measurements with ETSI and those from other Observatories \label{tb:exo_comparison}}
\tablehead{\colhead{Planet Name} & \colhead{Anderson-Darling} & \colhead{Anderson-Darling} & \colhead{Instrument} & \colhead{Telescope} & \colhead{References} \\
 & \colhead{p-value Combined} & \colhead{p-value Single} & & & }

\startdata
HAT-P-12~b & 0.52 & 0.33 & STIS & HST & \citet{Sing:2016} \\
           &      & 0.70 & MODS & LBT & \citet{Yan:2020}\\
\hline
HAT-P-32~b & 0.69 & 0.35 & GMOS   & Gemini & \citet{Gibson:2013}\\
           &      & 0.04 & OSIRIS & GTC    & \citet{Nortman:2016}\\
           &      & 0.16 &        & Multiple    & \citet{Mallonn:2016}\\
           &      & 0.60 & STIS   & HST    & \citet{Alam:2020}\\
\hline
HD 209458~b & 0.67 & -- &  STIS & HST & \citet{Sing:2016} \\
\hline
WASP-33~b & 0.24 & -- & OSIRIS & GTC & \citet{vonEsson:2019}\\
\hline
WASP-52~b & 0.33 & 0.05 & OSIRIS & GTC & \citet{Chen:2017}\\
          &      & 0.54 & ACAM & WHT & \citet{Louden:2017}\\
          &      & 0.61 & STIS & HST & \citet{Alam:2018}\\
\hline
WASP-69~b & 0.91 &  --   &  OSIRIS & GTC & \citet{Murgas:2020}\\
\hline
WASP-74~b & 0.38 &   --  & STIS & HST & \citet{Fu:2021} \\
\hline
WASP-103~b & 0.68 & 0.33 & FORS2 & VLT & \citet{Wilson:2020} \\
           &      & 0.79 & -- & DFOSC \& GROND & \citet{Southworth:2016} \\
\enddata

\end{deluxetable}

\clearpage

\begin{deluxetable}{cccccccccc}
\tabletypesize{\scriptsize}
\tablewidth{0pt}
\tablecaption{Parameter Values for TSM$_e$ Calculation \label{tb:metric}}
\tablehead{\colhead{Planet Name} & \colhead{TSM$_e$} & \colhead{Star T$_{eff}$} & \colhead{Star Radius}& \colhead{a} & \colhead{Planet T$_{eff}$} & \colhead{Planet Mass} & \colhead{$J$} & \colhead{$\delta_T$} & \colhead{Scale Factor (s)} \\
 &  & K & R$_{\odot}$ & AU & K & M$_J$ & & & }

\startdata
TrES-3~b & 264 & 5650 & 0.83 & 0.02 & 1643 & 1.91 & 11.015 & 0.027 & 17.63\\ 
WASP-69~b & 259 & 4700 & 0.86 & 0.05 & 988 & 0.29 & 8.032 & 0.018 & 1.77\\ 
TrES-1~b & 133 & 5230 & 0.85 & 0.04 & 1173 & 0.84 & 10.294 & 0.019 & 5.91\\ 
HAT-P-12~b & 121 & 4650 & 0.7 & 0.04 & 957 & 0.21 & 10.794 & 0.02 & 1.91\\ 
WASP-103~b & 119 & 6110 & 1.44 & 0.02 & 2510 & 1.49 & 11.1 & 0.012 & 9.22\\ 
HAT-P-44~b & 103 & 5295 & 0.95 & 0.05 & 1105 & 0.35 & 11.729 & 0.018 & 3.98\\ 
HAT-P-27~b & 91 & 5316 & 0.86 & 0.04 & 1189 & 0.62 & 10.626 & 0.015 & 4.2\\ 
HAT-P-17~b & 85 & 5246 & 0.87 & 0.09 & 794 & 0.58 & 9.017 & 0.015 & 2.6\\ 
WASP-77A~b & 80 & 5617 & 0.91 & 0.02 & 1691 & 1.67 & 8.766 & 0.019 & 2.63\\ 
HD209458~b & 73 & 6091 & 1.19 & 0.05 & 1477 & 0.73 & 6.591 & 0.014 & 0.5\\ 
WASP-52~b & 72 & 5000 & 0.79 & 0.03 & 1299 & 0.46 & 10.588 & 0.027 & 1.25\\ 
TrES-2~b & 64 & 5850 & 1.12 & 0.04 & 1580 & 1.49 & 10.232 & 0.016 & 4.17\\ 
WASP-33~b & 62 & 7430 & 1.44 & 0.02 & 2782 & 2.09 & 7.581 & 0.013 & 1.37\\ 
HAT-P-3~b & 35 & 5185 & 0.87 & 0.04 & 1185 & 0.65 & 9.936 & 0.012 & 1.57\\ 
XO-1~b & 28 & 5750 & 0.88 & 0.05 & 1173 & 0.83 & 9.939 & 0.018 & 1.09\\ 
HAT-P-32~b & 26 & 6001 & 1.37 & 0.03 & 1838 & 0.68 & 10.251 & 0.022 & 0.47\\ 
KELT-23A~b & 20 & 5899 & 1.0 & 0.03 & 1566 & 0.94 & 9.208 & 0.018 & 0.48\\ 
WASP-74~b & 16 & 5990 & 1.42 & 0.04 & 1810 & 0.72 & 8.548 & 0.01 & 0.36\\ 
CoRoT-2~b & 13 & 5625 & 0.91 & 0.03 & 1547 & 3.47 & 10.783 & 0.027 & 1.53\\ 
WASP-48~b & 10 & 5920 & 1.58 & 0.03 & 1943 & 0.8 & 10.627 & 0.01 & 0.72\\ 
KELT-9~b & 2 & 10170 & 2.36 & 0.03 & 4050 & 2.88 & 7.458 & 0.007 & 0.08\\ 
\enddata

\end{deluxetable}

\clearpage

\bibliographystyle{apj}
\bibliography{references}

\end{document}